\newcommand{\bq}{\begin{equation}}
\newcommand{\eq}{\end{equation}}
\newcommand{\bqa}{\begin{eqnarray}}
\newcommand{\eqa}{\end{eqnarray}}

\newcommand{\threev}[4]{\left(\begin{tabular}{c} $#1$ \\ $#2$ \\ $#3$ \end{tabular}\right)}

\newcommand{\f}{\varphi}
\newcommand{\h}{{1\over2}}

\documentclass[a4paper]{article}
\usepackage{epsfig}
\usepackage{axodraw}
\usepackage{graphicx}
\usepackage{amsmath}
\usepackage{makeidx}
\usepackage[top=1in, bottom=1in, left=1in, right=1in]{geometry}

\begin{document}

\begin{center}
\LARGE
The Path-Integral Approach to the $N=2$ Linear Sigma Model\\[30pt]
\large
E.N. Argyres\\[5pt]
Institute of Nuclear Physics, NCSR "Demokritos", Athens, Greece\\[20pt]
M.T.M. van Kessel\footnote{M.vanKessel@science.ru.nl} \hspace{50pt} R.H.P. Kleiss\footnote{R.Kleiss@science.ru.nl}\\[5pt]
IMAPP, FNWI, Radboud Universiteit Nijmegen, Nijmegen, The Netherlands\\[20pt]
December 3, 2008\\[50pt]
\end{center}

\section{Abstract}

In QFT the effective potential is an important tool to study symmetry breaking phenomena. It is known that, in some theories, the canonical approach and the path-integral approach yield different effective potentials. In this paper we investigate this for the Euclidean $N=2$ linear sigma model. Both the Green's functions and the effective potential will be computed in three different ways. The relative merits of the various approaches are discussed.

\newpage

\section{Introduction}

In the last couple of decades the Standard Model has become the generally accepted theory of fundamental physics. Many experiments have been performed and all results agreed with the Standard Model. Also all the particles that the Standard Model predicts have been detected in an actual experiment, except for one: the Higgs particle.

For this reason the Higgs sector of the Standard Model is very important and interesting. The Higgs mechanism was proposed in the 60's by Brout and Englert \cite{Englert}, Higgs \cite{Higgs, Higgs2} and Guralnik, Hagen and Kibble \cite{Guralnik} to give masses to the gauge bosons and the fermions, while keeping the theory renormalizable. The main feature of this Higgs mechanism is the mechanism of spontaneous symmetry breaking (SSB), which was introduced into quantum field theory by Nambu \cite{Nambu, NambuJona-Lasinio}, in analogy to the BCS theory of superconductivity.

A nice introduction to SSB and the Higgs mechanism can be found in a review article by Bernstein \cite{Bernstein}.

\subsection{Spontaneous Symmetry Breaking}

How does SSB work in quantum field theory, and what is it? The \emph{canonical} approach to SSB, which one finds in most textbooks (e.g.\ \cite{Peskin, Weinberg, Itzykson}), is as follows. One starts with a (bare) Lagrangian, obeying some symmetry in the fields (e.g. reflection or rotational symmetry), of which the bare (classical) potential has \emph{more} than one minimum. (The most common, and most important, example is the `Mexican hat' potential.) This means that the set of minima must also obey the symmetry, which means again that in any given minimum the fields cannot all be zero. Writing all fields into the single \emph{vector} $\f$ we have at the minima: $\f\neq0$. Therefore the classical lowest energy states, or vacua, are \emph{degenerate} and have a \emph{non-zero} field value, $\f_{\textrm{VAC}}\neq0$. In a quantum field theory the lowest energy state, or vacuum $|\textrm{VAC}\rangle$, should be calculated from the Schr\"odinger equation:
\bq
H|\textrm{VAC}\rangle = E_{\textrm{VAC}}|\textrm{VAC}\rangle \;.
\eq
Clearly, because of the very complicated form of the Hamiltonian $H$ in a quantum field theory, this equation can not be solved. Inspired by the classical minimum-energy states, one therefore \emph{postulates} that also the quantum vacuum is degenerate, and that:
\bq \label{vacpostulate}
\langle\textrm{VAC}|\f|\textrm{VAC}\rangle \neq 0 \;.
\eq
So there are \emph{multiple} vacuum states. But we can only live in one of these, and nature has chosen one of these vacuum states. Which one has been chosen, cannot be determined, and is therefore unimportant, because all theories built on one of these states have exactly the same physics.

This is called spontaneous symmetry breaking, i.e.\ the vacuum state of the theory does \emph{not} have the same symmetry as the Lagrangian. So the dynamics of the theory obey a certain symmetry, which is not respected by the vacuum state.

Having postulated (\ref{vacpostulate}) one can then derive, via the equations of motion, the Schwinger-Dyson equations and the Feynman rules, that this gives a mass-like term for all particles coupling to the (Higgs) field $\f$. The fluctuation in this (Higgs) field around the constant value it has in the chosen vacuum is the Higgs particle.

After this one can calculate all Green's functions of the theory. Also one can construct the 1PI Green's functions and sum them, in the appropriate way, to obtain the effective potential. As we shall see in section \ref{can} this effective potential comes out to be \emph{complex} and can be \emph{non-convex} in certain domains. This is the well known \emph{convexity problem}, i.e.\ the canonical perturbative calculation gives a non-convex effective potential, whereas general arguments show that this effective potential is \emph{convex}. 

In \cite{Symanzik, Iliopoulos} Symanzik and Iliopoulos et al.\ were the first to realize that the effective potential is always convex. A nice proof of this convexity property is given by Haymaker et al\ \cite{Haymaker}. (See also the PhD thesis of one of the authors \cite{vanKessel}.) Note that this proof is based on the path-integral formalism. The fact that there is a convexity problem was first realized by O'Raifeartaigh et al.\ \cite{ORaifeartaigh}. After this there were several attempts to modify the computation of the effective potential to find a proper, convex effective potential. These attempts can be found in \cite{Fujimoto, Callaway, Bender, Cooper, Hindmarsh, Rivers}. Indeed these attempt were successful, in all of these articles a convex, well-defined effective potential is found for several models. All these attempts come down to the same idea, to get a convex and well-defined effective potential one should take the path integral seriously and calculate from there. This means one should include \emph{all} minima of the classical potential in the calculation, i.e.\ do perturbation theory around \emph{each} of the minima and add the generating functionals around each of the minima to obtain the \emph{complete} generating functional. If one then computes the effective potential from this complete generating functional one finds the result to be convex and well-defined for all field values.

However, in this new (path-integral) approach, SSB is lost in the strict sense, i.e.\ all of the convex effective potentials that are calculated in the articles above have their minimum at zero for finite space-time volume. For infinite volume the bottom of the effective potentials becomes flat (Maxwell construction) and one is left with an infinite set of minima, living between the classical minima. What, then, is the true vacuum? Can one still determine what the vacuum is from these effective potentials? In \cite{Fujimoto} one can find a short remark about this. There the authors state that in the case of a non-convex classical potential, maybe the effective potential is \emph{not} the proper thing to look at to find the true vacuum. Or alternatively one might define SSB not as a non-zero vacuum expectation value, but as the sensitivity of the effective potential to small external sources. In this sense the new, convex effective potential is just as sensitive to a non-zero source as the old, non-convex effective potential.

However, besides these few vague remarks, no clear explanation is given as to what the path-integral approach means for the physics of the theory.

O'Raifeartaigh et al.\ \cite{Wipf}, inspired by \cite{Fukuda}, introduce a constraint effective potential. This constraint effective potential is calculated from a path integral, in which a constraint that keeps the space-time averaged field to a non-zero value is included. Simply because of the constraint, there is SSB in the strict sense now. However in the infinite volume limit the constraint effective potential converges to the convex effective potential again, leaving one again with a flat bottom of minima. Again it is unclear what this means for the physics. Also Ringwald et al.\ \cite{Ringwald} define a constraint effective potential, however now the constraint keeps the average of the field over a certain limited domain of space-time to a non-zero value. Again the constraint effective potential converges to the convex effective potential in the infinite volume limit. Nothing is said about the physics behind this theory.

Branchina et al.\ \cite{Branchina} do go into more details about the physics. Here they also include all minima of the path integral (no constraint) and find a flat bottom. Their approach is essentially based on the canonical formalism and they find explicitly the ground states in a Gaussian approximation. They find two pure Gaussian states, which means that all linear superpositions of these states are also ground states. These correspond to the flat bottom of the effective potential. They calculate the probability to be in one of these states. This probability is only non-zero for the pure Gaussian states. This is their interpretation of SSB. However, for the rest nothing is said about the physics that follows from this approach.

Weinberg et al.\ \cite{Wu} further analyze the complex, non-convex effective potential one finds when only including \emph{one} minimum (i.e.\ canonical approach). They define the vacuum states of the theory to be states that, of course, minimize the Hamiltonian, but are also localized around some field value. It appears that the imaginary part of the complex effective potential is related to the decay rate of the (unstable) vacuum states which are localized around a point between the classical minima.

Dannenberg \cite{Dannenberg} further analyzes and resolves the convexity problem. The point is that the convex effective potential, as calculated from the path integral, and the complex effective potential, as calculated in the canonical way (the sum of all 1PI diagrams), are simply not the same thing. In the path-integral approach one includes \emph{all} minima, in the canonical approach one includes only \emph{one} minimum. Although both ways are solutions to the same Schwinger-Dyson equations, they are \emph{not} equal. In this way it is completely understandable that the canonical approach gives a non-convex effective potential, even though one can prove from the path integral that the effective potential is convex. Both approaches are simply different and therefore give different results and physics.

Wiedemann \cite{Wiedemann} further analyzes what the non-convex complex effective potential and the convex effective potential tell one about the physics of the theory. It is shown that the flat section of the convex effective potential corresponds to the ground states of the theory. The complex effective potential gives one the boundaries of the flat section.

Having considered all of this literature one can conclude the following. The convexity problem is \emph{not} really a problem, it originates only because one compares two different things, at first thought to be the same. The canonical approach and the path-integral approach, although solutions to the same Schwinger-Dyson equation, seem to be \emph{different} in the case of a non-convex classical potential. So both approaches also give different results. This difference between the canonical and path-integral approach will be investigated in this paper in the case of the Euclidean $N=2$ linear sigma model. It is also this difference that might create some confusion in for example Peskin and Schroeder \cite{Peskin}. In their chapter 11 they first calculate the effective potential in the canonical approach and find it to be non-convex. Later they argue that the effective potential is always convex. They do not clearly explain how this convexity property relates to the non-convex result.

Taking the viewpoint of the canonical approach, one postulates a non-zero vacuum expectation value. This is completely self-consistent and one finds a spontaneously broken theory. One can define the effective potential as the sum of all 1PI graphs (with the appropriate factors) and one finds it to be non-convex and complex in certain regions. This does not matter however, since the proof that the effective potential is convex originates only in the path-integral approach, which is \emph{not} the same.

Taking the viewpoint of the path-integral approach one finds a convex effective potential, as can be proven on general grounds (within this approach). However, what the physics of this approach is, is unclear up to now. Also interesting is whether one can reproduce the physics as it is found in the canonical approach (with SSB and all) in this path-integral approach. Can one get the same Green's functions in this path-integral approach?

\subsection{Outline of this Paper}

In the articles mentioned above several links between results from the canonical approach and results from the path-integral approach are proven to exist, although both approaches do \emph{not} give the same results in general. This matter, specified to the case of the Euclidean $N=2$ linear sigma model ($N=2$ LSM), will be the main topic of this paper. This Euclidean version of the $N=2$ LSM is the second simplest model one can study to learn something about the differences between the canonical an path-integral approaches. (The Euclidean $N=1$ LSM is the simplest model one can study in this case, for calculations on this model we refer to the PhD thesis of one of the authors \cite{vanKessel} and to \cite{Fujimoto,Bender,Cooper}.)

In section \ref{can} we will present the canonical approach to the $N=2$ LSM. It should be emphasized here that the calculations in this section are similar to the standard calculations done in all textbooks, like \cite{Peskin}. This section is meant as an illustration of the convexity problem and an introduction for the later sections. The renormalized Green's functions will be computed and the counter terms will be fixed, such that we can use them later throughout the paper. The effective potential will also be calculated and shown to be complex where the classical potential is non-convex. Also it can become non-convex.

In section \ref{pathI} the path-integral approach to the $N=2$ LSM will be presented. We will compute the renormalized Green's functions by naively integrating over all minima of the action. This integration over minima is similar to what \cite{Fujimoto,Bender,Cooper} do in their case of the $N=1$ LSM. To our knowledge, this procedure has not been outlined in the high energy physics literature yet for a model with a continuous set of minima, like the $N=2$ LSM. Also an approximation to the effective potential will be found in this naive way. However, in the case of the $N=2$ LSM, which has a continuous set of minima (unlike the $N=1$ LSM), it is questionable whether the naive way of calculating here is correct.

In section \ref{pathII} we consider again the path-integral approach to the $N=2$ LSM. Now we do the calculations via the path integral in terms of polar variables. How this transformation to polar fields should be handled has been discussed in a previous paper by the authors \cite{vanKessel2}. By doing the calculations in this way we avoid the difficulties with the continuous set of minima that we saw in section \ref{pathI}. We will calculate the renormalized Green's functions and the effective potential. We will compare the results obtained here with the results from section \ref{pathI}, and finally discuss the physics of the path-integral approach to the $N=2$ LSM.

\subsection{The Euclidean $N=2$ LSM}

The model that we will be studying in this paper, the Euclidean $N=2$ LSM, has the following bare action:
\bq
S = \int d^dx \; \left( \h\left(\nabla\f_1(x)\right)^2 + \h\left(\nabla\f_2(x)\right)^2 - \h\mu\left(\f_1^2(x)+\f_2^2(x)\right) + {\lambda\over24}\left(\f_1^2(x)+\f_2^2(x)\right)^2 \right) \;.
\eq
Here $x$ denotes a $d$-vector containing all space-time coordinates 
\bq
x \equiv \threev{x_1}{\vdots}{x_d} \;,
\eq
and $\nabla$ is the $d$-vector
\bq
\nabla \equiv \threev{\partial/\partial x_1}{\vdots}{\partial/\partial x_d} \;.
\eq
Here we keep the dimension $d$ general, however $d$ is understood to be an integer. We shall \emph{not} employ the dimensional regularization scheme, but instead keep the regularization scheme general. In our explicit calculations this will come down to expressing everything in terms of standard integrals.

We take $\mu>0$, i.e.\ we consider the case that exhibits spontaneous symmetry breaking in the canonical approach.

\section{The Canonical Approach}\label{can}

In this section we will outline the canonical approach to the $N=2$ LSM. Our calculations will be much like those found in most textbooks, like e.g.\ Peskin and Schroeder \cite{Peskin}. 

\subsection{Green's Functions}

To compute the renormalized Green's functions of this theory we introduce renormalized quantities as follows:
\bqa
\f_{1/2}^{\mathrm{R}} &\equiv& {1\over\sqrt{Z}} \f_{1/2} \;, \quad Z \equiv 1 + \delta_Z \nonumber\\
\mu^{\mathrm{R}} &\equiv& \mu Z - \delta_{\mu} \nonumber\\
\lambda^{\mathrm{R}} &\equiv& \lambda Z^2 - \delta_{\lambda} \label{renquant}
\eqa
The action in terms of these renormalized quantities is (we will suppress the $R$-superscripts from now on):
\bqa
S &=& \int d^dx \; \bigg( \h\left(\nabla\f_1\right)^2 + \h\left(\nabla\f_2\right)^2 - \h\mu\left(\f_1^2+\f_2^2\right) + {\lambda\over24}\left(\f_1^2+\f_2^2\right)^2 + \nonumber\\[5pt]
& & \phantom{\int d^dx \; \bigg(} \h\delta_Z\left(\nabla\f_1\right)^2 + \h\delta_Z\left(\nabla\f_2\right)^2 - \h\delta_{\mu}\left(\f_1^2+\f_2^2\right) + {\delta_{\lambda}\over24}\left(\f_1^2+\f_2^2\right)^2 \bigg) \;.
\eqa
Now the classical action, i.e.\ the first line has its minima on the circle
\bq
\f_1^2+\f_2^2 = v^2 \;, \quad v \equiv \sqrt{{6\mu\over\lambda}} \;.
\eq
We choose
\bq
\f_1=v \;, \quad \f_2=0
\eq
as the true minimum in this canonical approach. Then define
\bq
\f_1 \equiv v+\eta_1 \;, \quad \f_2 \equiv \eta_2 \;.
\eq
In terms of these $\eta$-fields the action becomes:
\bqa
S &=& \int d^dx \; \bigg( \h\left(\nabla\eta_1\right)^2 + \h\left(\nabla\eta_2\right)^2 + \mu\eta_1^2 + {\mu\over v}\eta_1^3 + {\mu\over v}\eta_1\eta_2^2 + \nonumber\\[5pt]
& & \phantom{\int d^dx \; \bigg(} {\mu\over4v^2}\eta^4 + {\mu\over4v^2}\eta_2^2 + {\mu\over2v^2}\eta_1^2\eta_2^2 + \nonumber\\[5pt]
& & \phantom{\int d^dx \; \bigg(} \left(-v\delta_{\mu}+{1\over6}v^3\delta_{\lambda}\right)\eta_1 + \h\delta_Z\left(\nabla\eta_1\right)^2 + \left(-\h\delta_{\mu}+{1\over4}v^2\delta_{\lambda}\right)\eta_1^2 + \nonumber\\[5pt]
& & \phantom{\int d^dx \; \bigg(} \h\delta_Z\left(\nabla\eta_2\right)^2 + \left(-\h\delta_{\mu}+{1\over12}v^2\delta_{\lambda}\right)\eta_2^2 + {1\over6}v\delta_{\lambda}\eta_1^3 + {1\over6}v\delta_{\lambda}\eta_1\eta_2^2 + \nonumber\\[5pt]
& & \phantom{\int d^dx \; \bigg(} {1\over24}\delta_{\lambda}\eta_1^4 + {1\over24}\delta_{\lambda}\eta_2^4 + {1\over12}\delta_{\lambda}\eta_1^2\eta_2^2 \bigg) \;. \label{N2etaaction}
\eqa

Now define $\mu\equiv\h m^2$. Then the The Feynman rules are:
{\allowdisplaybreaks\bqa
\begin{picture}(100, 20)(0, 17)
\Line(20, 20)(80, 20)
\end{picture}
&\leftrightarrow& \frac{\hbar}{k^2 + m^2} \nonumber\\
\begin{picture}(100, 40)(0, 17)
\DashLine(20, 20)(80, 20){3}
\end{picture}
&\leftrightarrow& \frac{\hbar}{k^2} \nonumber\\
\begin{picture}(100, 40)(0, 17)
\Line(20, 20)(50, 20)
\Line(50, 20)(80, 0)
\Line(50, 20)(80, 40)
\end{picture}
&\leftrightarrow& -\frac{3m^2}{\hbar v} \nonumber\\
\begin{picture}(100, 40)(0, 17)
\Line(20, 20)(50, 20)
\DashLine(50, 20)(80, 0){3}
\DashLine(50, 20)(80, 40){3}
\end{picture}
&\leftrightarrow& -\frac{m^2}{\hbar v} \nonumber\\
\begin{picture}(100, 40)(0, 17)
\Line(20, 40)(80, 0)
\Line(20, 0)(80, 40)
\end{picture}
&\leftrightarrow& -\frac{3m^2}{\hbar v^2} \nonumber\\
\begin{picture}(100, 40)(0, 17)
\DashLine(20, 40)(80, 0){3}
\DashLine(20, 0)(80, 40){3}
\end{picture}
&\leftrightarrow& -\frac{3m^2}{\hbar v^2} \nonumber\\
\begin{picture}(100, 40)(0, 17)
\Line(20, 40)(80, 0)
\DashLine(20, 0)(80, 40){3}
\end{picture}
&\leftrightarrow& -\frac{m^2}{\hbar v^2} \nonumber\\
\begin{picture}(100, 40)(0, 17)
\Line(20, 20)(50, 20)
\Vertex(50,20){3}
\end{picture}
&\leftrightarrow& {v\over\hbar}\delta_{\mu} - {1\over6}{v^3\over\hbar}\delta_{\lambda} \nonumber\\
\begin{picture}(100, 40)(0, 17)
\Line(20, 20)(80, 20)
\Vertex(50,20){3}
\end{picture}
&\leftrightarrow& -{1\over\hbar}\delta_Z k^2 + {1\over\hbar}\delta_{\mu} - \h{v^2\over\hbar}\delta_{\lambda} \nonumber\\
\begin{picture}(100, 40)(0, 17)
\DashLine(20, 20)(80, 20){3}
\Vertex(50,20){3}
\end{picture}
&\leftrightarrow& -{1\over\hbar}\delta_Z k^2 + {1\over\hbar}\delta_{\mu} - {1\over6}{v^2\over\hbar}\delta_{\lambda} \nonumber\\
\begin{picture}(100, 40)(0, 17)
\Line(20, 20)(50, 20)
\Line(50, 20)(80, 0)
\Line(50, 20)(80, 40)
\Vertex(50,20){3}
\end{picture}
&\leftrightarrow& -{v\over\hbar}\delta_{\lambda} \nonumber\\
\begin{picture}(100, 40)(0, 17)
\Line(20, 20)(50, 20)
\DashLine(50, 20)(80, 0){3}
\DashLine(50, 20)(80, 40){3}
\Vertex(50,20){3}
\end{picture}
&\leftrightarrow& -{1\over3}{v\over\hbar}\delta_{\lambda} \nonumber\\
\begin{picture}(100, 40)(0, 17)
\Line(20, 40)(80, 0)
\Line(20, 0)(80, 40)
\Vertex(50,20){3}
\end{picture}
&\leftrightarrow& -{1\over\hbar}\delta_{\lambda} \nonumber\\
\begin{picture}(100, 40)(0, 17)
\DashLine(20, 40)(80, 0){3}
\DashLine(20, 0)(80, 40){3}
\Vertex(50,20){3}
\end{picture}
&\leftrightarrow& -{1\over\hbar}\delta_{\lambda} \nonumber\\
\begin{picture}(100, 40)(0, 17)
\Line(20, 40)(80, 0)
\DashLine(20, 0)(80, 40){3}
\Vertex(50,20){3}
\end{picture}
&\leftrightarrow& -{1\over3}{1\over\hbar}\delta_{\lambda}
\eqa}

\vspace{10pt}

Now we calculate the momentum space Green's functions of this theory, up to one loop. We shall write the results in terms of the standard integrals listed in appendix \ref{appstandint}.

Notice that standard integrals like $I(0,0,\ldots,0)$ are zero in the dimensional regularization scheme. We shall \emph{not} specify the regularization scheme here however, but instead keep everything general.
\bqa
\langle \tilde{\eta}_1 \rangle &=& \quad
\begin{picture}(50,20)(0,17)
\Line(0,20)(20,20)
\CArc(35,20)(15,-180,180)
\end{picture} \quad + \quad
\begin{picture}(50,20)(0,17)
\Line(0,20)(20,20)
\DashCArc(35,20)(15,-180,180){3}
\end{picture} \quad + \quad
\begin{picture}(50,20)(0,17)
\Line(0,20)(35,20)
\Vertex(35,20){3}
\end{picture} \nonumber\\[20pt]
&=& -{3\over2}{\hbar\over v} \; I(0,m) - \h{\hbar\over v} \; I(0,0) + {v\over m^2} \; \delta_{\mu}|_{\hbar} - {1\over6}{v^3\over m^2} \; \delta_{\lambda}|_{\hbar}
\eqa
\bq
\langle \tilde{\eta}_2 \rangle = 0
\eq
\bqa
\langle \tilde{\eta}_1\left(p\right) \tilde{\eta}_1\left(q\right) \rangle_{\mathrm{c}} &=& \quad
\begin{picture}(70,40)(0,17)
\Line(0,20)(70,20)
\end{picture} \quad + \quad
\begin{picture}(70,40)(0,17)
\Line(0,20)(20,20)
\CArc(35,20)(15,-180,180)
\Line(50,20)(70,20)
\end{picture} \quad + \quad
\begin{picture}(70,40)(0,17)
\Line(0,20)(20,20)
\DashCArc(35,20)(15,-180,180){3}
\Line(50,20)(70,20)
\end{picture} \quad + \nonumber\\
& & \quad \begin{picture}(70,40)(0,17)
\Line(0,5)(70,5)
\CArc(35,20)(15,-90,270)
\end{picture} \quad + \quad
\begin{picture}(70,40)(0,17)
\Line(0,5)(70,5)
\DashCArc(35,20)(15,-90,270){3}
\end{picture} \quad + \quad
\begin{picture}(70,40)(0,17)
\Line(0,5)(70,5)
\Line(35,5)(35,15)
\GCirc(35,25){10}{0.7}
\end{picture} \quad + \nonumber\\
& & \quad \begin{picture}(70,40)(0,17)
\Line(0,20)(70,20)
\Vertex(35,20){3}
\end{picture} \nonumber\\[25pt]
&=& \frac{\hbar}{p^2+m^2} + \frac{1}{\left(p^2+m^2\right)^2} \Bigg( {9\over2}{\hbar^2m^4\over v^2} \; I(0,m,p,m) + \h{\hbar^2m^4\over v^2} \; I(0,0,p,0) + \nonumber\\
& & \phantom{\frac{\hbar}{p^2+m^2} + \frac{1}{\left(p^2+m^2\right)^2} \Bigg(} 3{\hbar^2m^2\over v^2} \; I(0,m) + {\hbar^2m^2\over v^2} \; I(0,0) + \nonumber\\
& & \phantom{\frac{\hbar}{p^2+m^2} + \frac{1}{\left(p^2+m^2\right)^2} \Bigg(} -\hbar p^2 \; \delta_Z|_{\hbar} - 2\hbar \; \delta_{\mu}|_{\hbar} \Bigg) \label{eta1propcan}
\eqa
\bqa
\langle \tilde{\eta}_2\left(p\right) \tilde{\eta}_2\left(q\right) \rangle_{\mathrm{c}} &=& \quad
\begin{picture}(70,40)(0,17)
\DashLine(0,20)(70,20){3}
\end{picture} \quad + \quad
\begin{picture}(70,40)(0,17)
\DashLine(0,20)(20,20){3}
\CArc(35,20)(15,-180,0)
\DashCArc(35,20)(15,0,180){3}
\DashLine(50,20)(70,20){3}
\end{picture} \quad + \quad
\begin{picture}(70,40)(0,17)
\DashLine(0,5)(70,5){3}
\CArc(35,20)(15,-90,270)
\end{picture} \quad + \nonumber\\
& & \quad \begin{picture}(70,40)(0,17)
\DashLine(0,5)(70,5){3}
\DashCArc(35,20)(15,-90,270){3}
\end{picture} \quad + \quad
\begin{picture}(70,40)(0,17)
\DashLine(0,5)(70,5){3}
\Line(35,5)(35,15)
\GCirc(35,25){10}{0.7}
\end{picture} \quad + \quad
\begin{picture}(70,40)(0,17)
\DashLine(0,20)(70,20){3}
\Vertex(35,20){3}
\end{picture} \nonumber\\[25pt]
&=& \frac{\hbar}{p^2} + \frac{1}{\left(p^2\right)^2} \Bigg( {\hbar^2m^4\over v^2} \; I(0,0,p,m) + {\hbar^2m^2\over v^2} \; I(0,m) - {\hbar^2m^2\over v^2} \; I(0,0) + \nonumber\\
& & \phantom{\frac{\hbar}{p^2} + \frac{1}{\left(p^2\right)^2} \Bigg(} -\hbar p^2 \; \delta_Z|_{\hbar} \Bigg)
\eqa
\bq
\langle \tilde{\eta}_1\left(p\right) \tilde{\eta}_2\left(q\right) \rangle_{\mathrm{c}} = 0
\eq
{\allowdisplaybreaks\bqa
\langle \tilde{\eta}_1\left(q_1\right) \tilde{\eta}_1\left(q_2\right) \tilde{\eta}_1\left(q_3\right) \rangle_{\mathrm{1PI}} &=& \quad
\begin{picture}(50,40)(0,17)
\Line(0,20)(25,20)
\Line(25,20)(50,0)
\Line(25,20)(50,40)
\end{picture} \quad + \nonumber\\[5pt]
& & \quad \begin{picture}(70,40)(0,17)
\Line(0,20)(20,20)
\CArc(35,20)(15,-180,180)
\Line(48,27)(70,40)
\Line(48,13)(70,0)
\end{picture} \quad + \quad
\begin{picture}(70,40)(0,17)
\Line(0,20)(20,20)
\DashCArc(35,20)(15,-180,180){3}
\Line(48,27)(70,40)
\Line(48,13)(70,0)
\end{picture} \quad + \nonumber\\[5pt]
& & \quad \begin{picture}(70,40)(0,17)
\Line(0,20)(20,20)
\CArc(35,20)(15,-180,180)
\Line(50,20)(70,40)
\Line(50,20)(70,0)
\end{picture} \quad + \quad
\begin{picture}(70,40)(0,17)
\Line(0,20)(20,20)
\DashCArc(35,20)(15,-180,180){3}
\Line(50,20)(70,40)
\Line(50,20)(70,0)
\end{picture} \quad + \nonumber\\[5pt]
& & \quad \begin{picture}(70,40)(0,17)
\Line(0,5)(70,5)
\CArc(35,20)(15,-180,180)
\Line(48,27)(70,40)
\end{picture} \quad + \quad
\begin{picture}(70,40)(0,17)
\Line(0,5)(70,5)
\DashCArc(35,20)(15,-180,180){3}
\Line(48,27)(70,40)
\end{picture} \quad + \nonumber\\[5pt]
& & \quad \begin{picture}(70,40)(0,17)
\Line(0,35)(70,35)
\CArc(35,20)(15,-180,180)
\Line(48,13)(70,0)
\end{picture} \quad + \quad 
\begin{picture}(70,40)(0,17)
\Line(0,35)(70,35)
\DashCArc(35,20)(15,-180,180){3}
\Line(48,13)(70,0)
\end{picture} \quad + \nonumber\\[5pt]
& & \quad \begin{picture}(50,40)(0,17)
\Line(0,20)(25,20)
\Line(25,20)(50,0)
\Line(25,20)(50,40)
\Vertex(25,20){3}
\end{picture} \nonumber\\[25pt]
&=& -{3m^2\over\hbar v} - {27m^6\over v^3} \; I(0,m,q_1,m,-q_3,m) \nonumber\\
& & -{m^6\over v^3} \; I(0,0,q_1,0,-q_3,0) \nonumber\\
& & +{9m^4\over2v^3} \; \left( I(0,m,q_1,m) + I(0,m,q_2,m) + I(0,m,q_3,m) \right) \nonumber\\
& & +{m^4\over2v^3} \; \left( I(0,0,q_1,0) + I(0,0,q_2,0) + I(0,0,q_3,0) \right) \nonumber\\
& & -{v\over\hbar} \; \delta_{\lambda}|_{\hbar}
\eqa
\bq
\langle \tilde{\eta}_1\left(q_1\right) \tilde{\eta}_1\left(q_2\right) \tilde{\eta}_2\left(q_3\right) \rangle_{\mathrm{1PI}} = 0
\eq
\bqa
\langle \tilde{\eta}_1\left(q_1\right) \tilde{\eta}_2\left(q_2\right) \tilde{\eta}_2\left(q_3\right) \rangle_{\mathrm{1PI}} &=& \quad
\begin{picture}(50,40)(0,17)
\Line(0,20)(25,20)
\DashLine(25,20)(50,0){3}
\DashLine(25,20)(50,40){3}
\end{picture} \quad + \nonumber\\[5pt]
& & \quad \begin{picture}(70,40)(0,17)
\Line(0,20)(20,20)
\DashCArc(35,20)(15,-30,30){3}
\CArc(35,20)(15,30,-30)
\DashLine(48,27)(70,40){3}
\DashLine(48,13)(70,0){3}
\end{picture} \quad + \quad
\begin{picture}(70,40)(0,17)
\Line(0,20)(20,20)
\DashCArc(35,20)(15,30,-30){3}
\CArc(35,20)(15,-30,30)
\DashLine(48,27)(70,40){3}
\DashLine(48,13)(70,0){3}
\end{picture} \quad + \nonumber\\[5pt]
& & \quad \begin{picture}(70,40)(0,17)
\Line(0,20)(20,20)
\CArc(35,20)(15,-180,180)
\DashLine(50,20)(70,40){3}
\DashLine(50,20)(70,0){3}
\end{picture} \quad + \quad
\begin{picture}(70,40)(0,17)
\Line(0,20)(20,20)
\DashCArc(35,20)(15,-180,180){3}
\DashLine(50,20)(70,40){3}
\DashLine(50,20)(70,0){3}
\end{picture} \quad + \nonumber\\[5pt]
& & \quad \begin{picture}(70,40)(0,17)
\Line(0,5)(35,5)
\DashLine(35,5)(70,5){3}
\CArc(35,20)(15,30,270)
\DashCArc(35,20)(15,-90,30){3}
\DashLine(48,27)(70,40){3}
\end{picture} \quad + \quad
\begin{picture}(70,40)(0,17)
\Line(0,35)(35,35)
\DashLine(35,35)(70,35){3}
\CArc(35,20)(15,90,-30)
\DashCArc(35,20)(15,-30,90){3}
\DashLine(48,13)(70,0){3}
\end{picture} \quad + \nonumber\\[5pt]
& & \quad \begin{picture}(50,40)(0,17)
\Line(0,20)(25,20)
\DashLine(25,20)(50,0){3}
\DashLine(25,20)(50,40){3}
\Vertex(25,20){3}
\end{picture} \nonumber\\[25pt]
&=& -{m^2\over\hbar v} \nonumber\\
& & -{3m^6\over v^3} \; I(0,m,q_1,m,-q_3,0) - {m^6\over v^3} \; I(0,0,q_1,0,-q_3,m) \nonumber\\
& & +{3m^4\over2v^3} \; I(0,m,q_1,m) + {3m^4\over2v^3}\; I(0,0,q_1,0) \nonumber\\
& & +{m^4\over v^3} \; I(0,m,q_2,0) + {m^4\over v^3} \; I(0,m,q_3,0) \nonumber\\
& & -{v\over3\hbar} \; \delta_{\lambda}|_{\hbar}
\eqa
\bq
\langle \tilde{\eta}_2\left(q_1\right) \tilde{\eta}_2\left(q_2\right) \tilde{\eta}_2\left(q_3\right) \rangle_{\mathrm{1PI}} = 0
\eq
\bqa
\langle \tilde{\eta}_1\left(q_1\right) \cdots \tilde{\eta}_1\left(q_4\right) \rangle_{\mathrm{1PI}} &=& \quad
\begin{picture}(40,40)(0,17)
\Line(0,0)(40,40)
\Line(0,40)(40,0)
\end{picture} \quad + \quad
\begin{picture}(70,40)(0,17)
\Line(0,40)(22,27)
\Line(0,0)(22,13)
\CArc(35,20)(15,-180,180)
\Line(48,27)(70,40)
\Line(48,13)(70,0)
\end{picture} \quad + \quad \textrm{2 perm's} \quad + \nonumber\\[5pt]
& & \quad \begin{picture}(70,40)(0,17)
\Line(0,40)(22,27)
\Line(0,0)(22,13)
\DashCArc(35,20)(15,-180,180){3}
\Line(48,27)(70,40)
\Line(48,13)(70,0)
\end{picture} \quad + \quad \textrm{2 perm's} \quad + \nonumber\\[5pt]
& & \quad \begin{picture}(70,40)(0,17)
\Line(0,40)(20,20)
\Line(0,0)(20,20)
\CArc(35,20)(15,-180,180)
\Line(48,27)(70,40)
\Line(48,13)(70,0)
\end{picture} \quad + \quad \textrm{5 perm's} \quad + \nonumber\\[5pt]
& & \quad \begin{picture}(70,40)(0,17)
\Line(0,40)(20,20)
\Line(0,0)(20,20)
\DashCArc(35,20)(15,-180,180){3}
\Line(48,27)(70,40)
\Line(48,13)(70,0)
\end{picture} \quad + \quad \textrm{5 perm's} \quad + \nonumber\\[5pt]
& & \quad \begin{picture}(70,40)(0,17)
\Line(0,40)(20,20)
\Line(0,0)(20,20)
\CArc(35,20)(15,-180,180)
\Line(50,20)(70,40)
\Line(50,20)(70,0)
\end{picture} \quad + \quad \textrm{2 perm's} \quad + \nonumber\\[5pt]
& & \quad \begin{picture}(70,40)(0,17)
\Line(0,40)(20,20)
\Line(0,0)(20,20)
\DashCArc(35,20)(15,-180,180){3}
\Line(50,20)(70,40)
\Line(50,20)(70,0)
\end{picture} \quad + \quad \textrm{2 perm's} \quad + \quad
\begin{picture}(40,40)(0,17)
\Line(0,0)(40,40)
\Line(0,40)(40,0)
\Vertex(20,20){3}
\end{picture} \nonumber\\[25pt]
&=& -{3m^2\over\hbar v^2} \nonumber\\
& & +{81m^8\over v^4} \; I(0,m,q_1,m,q_1+q_3,m,-q_2,m) + \textrm{2 perm's} \nonumber\\
& & +{m^8\over v^4} \; I(0,0,q_1,0,q_1+q_3,0,-q_2,0) + \textrm{2 perm's} \nonumber\\
& & -{27m^6\over v^4} \; I(0,m,q_3,m,q_3+q_4,m) + \textrm{5 perm's} \nonumber\\
& & -{m^6\over v^4} \; I(0,0,q_3,0,q_3+q_4,0) + \textrm{5 perm's} \nonumber\\
& & +{9m^4\over2v^4} \; I(0,m,q_3+q_4,m) + \textrm{2 perm's} \nonumber\\
& & +{m^4\over2v^4} \; I(0,0,q_3+q_4,0) + \textrm{2 perm's} \nonumber\\
& & -{1\over\hbar} \; \delta_{\lambda}|_{\hbar}
\eqa
\bq
\langle \tilde{\eta}_1\left(q_1\right) \tilde{\eta}_1\left(q_2\right) \tilde{\eta}_1\left(q_3\right) \tilde{\eta}_2\left(q_4\right) \rangle_{\mathrm{1PI}} = 0
\eq
\bqa
\langle \tilde{\eta_1}\left(q_1\right) \tilde{\eta}_1\left(q_2\right) \tilde{\eta}_2\left(q_3\right) \tilde{\eta}_2\left(q_4\right) \rangle_{\mathrm{1PI}} &=& \quad
\begin{picture}(40,40)(0,17)
\Line(0,0)(20,20)
\Line(0,40)(20,20)
\DashLine(20,20)(40,40){3}
\DashLine(20,20)(40,0){3}
\end{picture} \quad + \quad
\begin{picture}(70,40)(0,17)
\Line(0,40)(22,27)
\Line(0,0)(22,13)
\CArc(35,20)(15,30,-30)
\DashCArc(35,20)(15,-30,30){3}
\DashLine(48,27)(70,40){3}
\DashLine(48,13)(70,0){3}
\end{picture} \quad + \quad \textrm{1 perm} \quad + \nonumber\\[5pt]
& & \quad \begin{picture}(70,40)(0,17)
\Line(0,40)(22,27)
\Line(0,0)(22,13)
\CArc(35,20)(15,-30,30)
\DashCArc(35,20)(15,30,-30){3}
\DashLine(48,27)(70,40){3}
\DashLine(48,13)(70,0){3}
\end{picture} \quad + \quad \textrm{1 perm} \quad + \nonumber\\[5pt]
& & \quad \begin{picture}(70,40)(0,17)
\Line(0,40)(20,20)
\Line(0,0)(20,20)
\CArc(35,20)(15,30,-30)
\DashCArc(35,20)(15,-30,30){3}
\DashLine(48,27)(70,40){3}
\DashLine(48,13)(70,0){3}
\end{picture} \quad + \quad
\begin{picture}(70,40)(0,17)
\Line(0,40)(20,20)
\Line(0,0)(20,20)
\CArc(35,20)(15,-30,30)
\DashCArc(35,20)(15,30,-30){3}
\DashLine(48,27)(70,40){3}
\DashLine(48,13)(70,0){3}
\end{picture} \quad + \nonumber\\[5pt]
& & \quad \begin{picture}(70,40)(0,17)
\Line(0,40)(22,27)
\Line(0,0)(22,13)
\CArc(35,20)(15,-180,180)
\DashLine(50,20)(70,40){3}
\DashLine(50,20)(70,0){3}
\end{picture} \quad + \quad
\begin{picture}(70,40)(0,17)
\Line(0,40)(22,27)
\Line(0,0)(22,13)
\DashCArc(35,20)(15,-180,180){3}
\DashLine(50,20)(70,40){3}
\DashLine(50,20)(70,0){3}
\end{picture} \quad + \nonumber\\[5pt]
& & \quad \begin{picture}(70,40)(0,17)
\Line(0,40)(35,35)
\DashLine(35,35)(70,40){3}
\CArc(35,20)(15,90,-30)
\DashCArc(35,20)(15,-30,90){3}
\Line(0,0)(22,13)
\DashLine(48,13)(70,0){3}
\end{picture} \quad + \quad \textrm{3 perm's} \quad + \nonumber\\[5pt]
& & \quad \begin{picture}(70,40)(0,17)
\Line(0,40)(35,35)
\DashLine(35,35)(70,40){3}
\CArc(35,20)(15,-30,90)
\DashCArc(35,20)(15,90,-30){3}
\Line(0,0)(22,13)
\DashLine(48,13)(70,0){3}
\end{picture} \quad + \quad \textrm{3 perm's} \quad + \nonumber\\[5pt]
& & \quad \begin{picture}(70,40)(0,17)
\Line(0,40)(20,20)
\Line(0,0)(20,20)
\CArc(35,20)(15,-180,180)
\DashLine(50,20)(70,40){3}
\DashLine(50,20)(70,0){3}
\end{picture} \quad + \quad
\begin{picture}(70,40)(0,17)
\Line(0,40)(20,20)
\Line(0,0)(20,20)
\DashCArc(35,20)(15,-180,180){3}
\DashLine(50,20)(70,40){3}
\DashLine(50,20)(70,0){3}
\end{picture} \quad + \nonumber\\[5pt]
& & \quad \begin{picture}(70,40)(0,17)
\Line(0,40)(35,35)
\DashLine(35,35)(70,40){3}
\CArc(35,20)(15,90,-90)
\DashCArc(35,20)(15,-90,90){3}
\Line(0,0)(35,5)
\DashLine(35,5)(70,0){3}
\end{picture} \quad + \textrm{1 perm} \quad + \nonumber\\[5pt]
& & \quad \begin{picture}(40,40)(0,17)
\Line(0,0)(20,20)
\Line(0,40)(20,20)
\DashLine(20,20)(40,40){3}
\DashLine(20,20)(40,0){3}
\Vertex(20,20){3}
\end{picture} \nonumber\\[25pt]
&=& -{m^2\over\hbar v^2} \nonumber\\
& & +{9m^8\over v^4} \; I(0,m,q_1,m,q_1+q_3,0,-q_2,m) + \textrm{1 perm} \nonumber\\
& & +{m^8\over v^4} \; I(0,0,q_1,0,q_1+q_3,m,-q_2,0) + \textrm{1 perm} \nonumber\\
& & -{9m^6\over v^4} \; I(0,m,q_3,0,q_3+q_4,m) \nonumber\\
& & -{m^6\over v^4} \; I(0,0,q_3,m,q_3+q_4,0) \nonumber\\
& & -{9m^6\over v^4} \; I(0,m,q_1,m,-q_2,m) \nonumber\\
& & -{3m^6\over v^4} \; I(0,0,q_1,0,-q_2,0) \nonumber\\
& & -{3m^6\over v^4} \; I(0,0,q_4,m,q_2+q_4,m) + \textrm{3 perm's} \nonumber\\
& & -{m^6\over v^4} \; I(0,m,q_4,0,q_2+q_4,0) + \textrm{3 perm's} \nonumber\\
& & +{3m^4\over2v^4} \; I(0,m,q_3+q_4,m) \nonumber\\
& & +{3m^4\over2v^4} \; I(0,0,q_3+q_4,0) \nonumber\\
& & +{m^4\over v^4} \; I(0,0,q_2+q_4,m) + \textrm{1 perm} \nonumber\\
& & -{1\over3\hbar} \; \delta_{\lambda}|_{\hbar}
\eqa
\bq
\langle \tilde{\eta}_1\left(q_1\right) \tilde{\eta}_2\left(q_2\right) \ldots \tilde{\eta}_2\left(q_4\right) \rangle_{\mathrm{1PI}} = 0
\eq
\bqa
\langle \tilde{\eta}_2\left(q_1\right) \ldots \tilde{\eta}_2\left(q_4\right) \rangle_{\mathrm{1PI}} &=& \quad
\begin{picture}(40,40)(0,17)
\DashLine(0,0)(40,40){3}
\DashLine(0,40)(40,0){3}
\end{picture} \quad + \quad
\begin{picture}(70,40)(0,17)
\DashLine(0,40)(22,27){3}
\DashLine(0,0)(22,13){3}
\CArc(35,20)(15,30,150)
\DashCArc(35,20)(15,150,210){3}
\CArc(35,20)(15,210,-30)
\DashCArc(35,20)(15,-30,30){3}
\DashLine(48,27)(70,40){3}
\DashLine(48,13)(70,0){3}
\end{picture} \quad + \quad \textrm{2 perm's} \quad + \nonumber\\[5pt]
& & \quad \begin{picture}(70,40)(0,17)
\DashLine(0,40)(22,27){3}
\DashLine(0,0)(22,13){3}
\CArc(35,20)(15,-30,30)
\DashCArc(35,20)(15,30,150){3}
\CArc(35,20)(15,150,210)
\DashCArc(35,20)(15,210,-30){3}
\DashLine(48,27)(70,40){3}
\DashLine(48,13)(70,0){3}
\end{picture} \quad + \quad \textrm{2 perm's} \quad + \nonumber\\[5pt]
& & \quad \begin{picture}(70,40)(0,17)
\DashLine(0,40)(20,20){3}
\DashLine(0,0)(20,20){3}
\CArc(35,20)(15,30,-30)
\DashCArc(35,20)(15,-30,30){3}
\DashLine(48,27)(70,40){3}
\DashLine(48,13)(70,0){3}
\end{picture} \quad + \quad \textrm{5 perm's} \quad + \nonumber\\[5pt]
& & \quad \begin{picture}(70,40)(0,17)
\DashLine(0,40)(20,20){3}
\DashLine(0,0)(20,20){3}
\CArc(35,20)(15,-30,30)
\DashCArc(35,20)(15,30,-30){3}
\DashLine(48,27)(70,40){3}
\DashLine(48,13)(70,0){3}
\end{picture} \quad + \quad \textrm{5 perm's} \quad + \nonumber\\[5pt]
& & \quad \begin{picture}(70,40)(0,17)
\DashLine(0,40)(20,20){3}
\DashLine(0,0)(20,20){3}
\CArc(35,20)(15,-180,180)
\DashLine(50,20)(70,40){3}
\DashLine(50,20)(70,0){3}
\end{picture} \quad + \quad \textrm{2 perm's} \quad + \nonumber\\[5pt]
& & \quad \begin{picture}(70,40)(0,17)
\DashLine(0,40)(20,20){3}
\DashLine(0,0)(20,20){3}
\DashCArc(35,20)(15,-180,180){3}
\DashLine(50,20)(70,40){3}
\DashLine(50,20)(70,0){3}
\end{picture} \quad + \quad \textrm{2 perm's} \quad + \nonumber\\[5pt]
& & \quad \begin{picture}(40,40)(0,17)
\Line(0,0)(20,20)
\Line(0,40)(20,20)
\DashLine(20,20)(40,40){3}
\DashLine(20,20)(40,0){3}
\Vertex(20,20){3}
\end{picture} \nonumber\\[25pt]
&=& -{3m^2\over\hbar v^2} \nonumber\\
& & +{m^8\over v^4} \; I(0,0,q_1,m,q_1+q_3,0,-q_2,m) + \textrm{2 perm's} \nonumber\\
& & +{m^8\over v^4} \; I(0,m,q_1,0,q_1+q_3,m,-q_2,0) + \textrm{2 perm's} \nonumber\\
& & -{m^6\over v^4} \; I(0,m,q_3,0,q_3+q_4,m) + \textrm{5 perm's} \nonumber\\
& & -{3m^6\over v^4} \; I(0,0,q_3,m,q_3+q_4,0) + \textrm{5 perm's} \nonumber\\
& & +{m^4\over2v^4} \; I(0,m,q_3+q_4,m) + \textrm{2 perm's} \nonumber\\
& & +{9m^4\over2v^4} \; I(0,0,q_3+q_4,0) + \textrm{2 perm's} \nonumber\\
& & -{1\over\hbar} \; \delta_{\lambda}|_{\hbar}
\eqa}

Now our theory contains three free parameters, $Z$, $\mu$ and $\lambda$, which have to be fixed by three renormalization conditions. We could try to use the following renormalization conditions:
\bqa
\begin{picture}(50,20)(0,17)
\Line(0,20)(20,20)
\GCirc(35,20){15}{0.5}
\end{picture} \quad &=& 0 \nonumber\\
\textrm{Res} \quad
\begin{picture}(70,40)(0,17)
\Line(0,20)(70,20)
\GCirc(35,20){15}{0.5}
\end{picture} \quad &=& \hbar \nonumber\\[5pt]
\begin{picture}(70,40)(0,17)
\Line(0,40)(70,0)
\Line(0,0)(70,40)
\GCirc(35,20){15}{0.5}
\Text(35,20)[cc]{1PI}
\end{picture} \quad &=& -{\lambda\over\hbar} \quad \textrm{at $q_1=\ldots=q_4=0$} \label{rencond2}
\eqa\\
This is \emph{not} a good idea however, because some of our amplitudes are singular at zero incoming momentum. These singularities are of course caused by loops with the Goldstone boson. If we would set the 4-point 1PI amplitude to $-\lambda/\hbar$ at zero external momenta we would be absorbing infrared divergences, which only occur for very specific external momenta, in the counter terms. The most straightforward thing to do now is change the renormalization point. However, this will complicate the calculations greatly.

What we shall do is just remove these infrared divergences from our counter term $\delta_{\lambda}$ by hand.

This is somewhat similar to what is done in Peskin and Schroeder \cite{Peskin}, there they work in the dimensional regularization scheme, in which the infrared divergences are invisible anyway. (Their renormalization point is at $s=4m^2, \; u=t=0$ however, but also at this point there occur IR divergences.)

If we strictly use the conditions (\ref{rencond2}) the counter terms become, up to order $\hbar$:
\bqa
\delta_{\mu}|_{\hbar} &=& {3\over2}{\hbar m^2\over v^2} \; I(0,m) + \h{\hbar m^2\over v^2} \; I(0,0) + {1\over6}v^2 \; \delta_{\lambda}|_{\hbar} \nonumber\\
\delta_{\lambda}|_{\hbar} &=& {27\over2}{\hbar m^4\over v^4} \; I(0,m,0,m) + {3\over2}{\hbar m^4\over v^4} \; I(0,0,0,0) + \nonumber\\
& & -162{\hbar m^6\over v^4} \; I(0,m,0,m,0,m) - 6{\hbar m^6\over v^4} \; I(0,0,0,0,0,0) + \nonumber\\
& & 243{\hbar m^8\over v^4} \; I(0,m,0,m,0,m,0,m) + 3{\hbar m^8\over v^4} \; I(0,0,0,0,0,0,0,0) \nonumber\\
\delta_Z|_{\hbar} &=& {9\over2}{\hbar m^4\over v^2} \; \left. {d\over dp^2} I(0,m,p,m) \right|_{p^2=-m_{\mathrm{ph,1}}^2} + \h{\hbar m^4\over v^2} \; \left. {d\over dp^2} I(0,0,p,0) \right|_{p^2=-m_{\mathrm{ph,1}}^2}
\eqa
Now we see that the second, fourth and sixth term in $\delta_{\lambda}|_{\hbar}$ contain infrared divergences, which we should \emph{not} include. The easiest thing to do is introduce a mass in these terms, such that the infrared divergences are regularized. We shall just use $m$ for this mass, to keep the calculation as simple as possible. After this manual procedure the counter terms are:
\bqa
\delta_{\mu}|_{\hbar} &=& {3\over2}{\hbar m^2\over v^2} \; I(0,m) + \h{\hbar m^2\over v^2} \; I(0,0) + {5\over2}{\hbar m^4\over v^2} \; I(0,m,0,m) + \nonumber\\
& & -28{\hbar m^6\over v^2} \; I(0,m,0,m,0,m) + 41{\hbar m^8\over v^2} \; I(0,m,0,m,0,m,0,m) \nonumber\\
\delta_{\lambda}|_{\hbar} &=& 15{\hbar m^4\over v^4} \; I(0,m,0,m) - 168{\hbar m^6\over v^4} \; I(0,m,0,m,0,m) + \nonumber\\
& & 246{\hbar m^8\over v^4} \; I(0,m,0,m,0,m,0,m) \nonumber\\
\delta_Z|_{\hbar} &=& {9\over2}{\hbar m^4\over v^2} \; \left. {d\over dp^2} I(0,m,p,m) \right|_{p^2=-m_{\mathrm{ph,1}}^2} + \h{\hbar m^4\over v^2} \; \left. {d\over dp^2} I(0,0,p,0) \right|_{p^2=-m_{\mathrm{ph,1}}^2} \label{counttermsN2}
\eqa

The physical masses of the $\eta_1$- and $\eta_2$-particle, $m_{\mathrm{ph,1}}$ and $m_{\mathrm{ph,2}}$, can now be calculated from the Dyson summed propagators. The Dyson summed $\eta_1$-propagator is
\bq \label{eta1prop}
\frac{\hbar}{p^2+m^2-\hbar A_1(p^2)}
\eq
with
\bqa
A_1(p^2) &=& {9\over2}{m^4\over v^2} \; I(0,m,p,m) - {9\over2}{m^4\over v^2} p^2 \; \left. {d\over dp^2} I(0,m,p,m) \right|_{p^2=-m_{\mathrm{ph,1}}^2} + \nonumber\\
& & \h{m^4\over v^2} \; I(0,0,p,0) - \h{m^4\over v^2} p^2 \; \left. {d\over dp^2} I(0,0,p,0) \right|_{p^2=-m_{\mathrm{ph,1}}^2} + \nonumber\\
& & -5{m^4\over v^2} \; I(0,m,0,m) + 56{m^6\over v^2} \; I(0,m,0,m,0,m) + \nonumber\\
& & -82{m^8\over v^2} \; I(0,m,0,m,0,m,0,m) \;. \label{A1canon}
\eqa
The location of the pole of (\ref{eta1prop}) gives $-m_{\mathrm{ph,1}}^2$. Up to order $\hbar$ we can easily find this pole:
\bqa
m_{\mathrm{ph,1}}^2 &=& m^2 + 5{\hbar m^4\over v^2} \; I(0,m,0,m) + \nonumber\\
& & -{9\over2}{\hbar m^4\over v^2} \; \left. I(0,m,p,m) \right|_{p^2=-m^2} - \h{\hbar m^4\over v^2} \; \left. I(0,0,p,0) \right|_{p^2=-m^2} + \nonumber\\
& & -{9\over2}{\hbar m^6\over v^2} \; \left. {d\over dp^2} I(0,m,p,m) \right|_{p^2=-m^2} - \h{\hbar m^6\over v^2} \; \left. {d\over dp^2} I(0,0,p,0) \right|_{p^2=-m^2} + \nonumber\\
& & -56{\hbar m^6\over v^2} \; I(0,m,0,m,0,m) + 82{\hbar m^8\over v^2} \; I(0,m,0,m,0,m,0,m) \;.
\eqa
For $d\leq4$ this $m_{\mathrm{ph,1}}$ is finite, for $d>4$ it is not, which shows that the LSM is non-renormalizable for $d>4$.

Likewise we can obtain the physical mass of the $\eta_2$-particle $m_{\mathrm{ph,2}}$. The Dyson summed $\eta_2$-propagator is:
\bq \label{eta2prop}
\frac{\hbar}{p^2-\hbar A_2(p^2)}
\eq
with
\bqa
A_2(p^2) &=& {m^4\over v^2} \; I(0,0,p,m) + {m^2\over v^2} \; I(0,m) - {m^2\over v^2} \; I(0,0) + \nonumber\\
& & -{9\over2}{m^4\over v^2} p^2 \; \left. {d\over dp^2} I(0,m,p,m) \right|_{p^2=-m_{\mathrm{ph,1}}^2} - \h{m^4\over v^2} p^2 \; \left. {d\over dp^2} I(0,0,p,0) \right|_{p^2=-m_{\mathrm{ph,1}}^2} \;. \label{A2canon}
\eqa
Again the pole of (\ref{eta2prop}) is easily found up to order $\hbar$:
\bq \label{mph2}
m_{\mathrm{ph,2}}^2 = -{\hbar m^4\over v^2} \; I(0,0,0,m) - {\hbar m^2\over v^2} \; I(0,m) + {\hbar m^2\over v^2} \; I(0,0) = 0
\eq
This is an illustration of the Goldstone theorem, which states that in the case of spontaneous symmetry breaking the mass of the Goldstone boson remains zero at \emph{all} orders.

Actually things are a bit trickier than they look here. The above result for $m_{\mathrm{ph,2}}$ seems to hold for all dimensions below 5, where the theory is renormalizable. This would mean that also for $d=1$, where we know that \emph{no} spontaneous symmetry breaking can occur, $m_{\mathrm{ph,2}}$ would remain zero up to order $\hbar$. This is \emph{not} true in general. In general (\ref{mph2}) is wrong for $d=1$ because we put the momentum $p$, flowing through the propagator, to zero \emph{before} we have done the loop integral. Actually we have to compute the integral and only then put $p$ to zero. The two operations do not commute. In case of the $N=2$ LSM it happens to be that (\ref{mph2}) is correct after all, the problem with setting $p$ to zero before doing the loop integrals only shows up in 2-loop integrals. One can explicitly verify that at 2-loop order $m_{\mathrm{ph,2}}$ is no longer zero for $d=1$. For $d>2$, (\ref{mph2}) is \emph{always} correct however, which is in complete agreement with the Goldstone theorem. (Remember that $d=2$ is a special case, see Coleman \cite{Coleman} and Coleman et al.\ \cite{Jackiw}.

\subsection{The Effective Potential}

Now we want to calculate the effective potential. We will use the vacuum-graph formula. However this calculation will be involved because we have \emph{two} types of lines, which complicates how we connect the lines inside the loop.

To deal with this complication we consider the same vertex, of which a different set of legs is going to be part of the loop, as \emph{different}. In this way each 1-loop diagram is characterized by 8 numbers, each denoting the number of a certain type of vertices in the diagram. These numbers are defined as follows:
\vspace{-17pt}
\bq
\begin{array}{lll}
\begin{picture}(100, 40)(0, 17)
\Line(20, 20)(50, 20)
\Line(50, 20)(80, 0)
\Line(50, 20)(80, 40)
\end{picture}
\leftrightarrow n_3 &
\begin{picture}(100, 40)(0, 17)
\Line(20, 20)(50, 20)
\DashLine(50, 20)(80, 0){3}
\DashLine(50, 20)(80, 40){3}
\end{picture}
\leftrightarrow m_3 &
\begin{picture}(100, 40)(0, 17)
\DashLine(20, 20)(50, 20){3}
\DashLine(50, 20)(80, 0){3}
\Line(50, 20)(80, 40)
\end{picture}
\leftrightarrow q_3 \\[10pt]
\begin{picture}(100, 40)(0, 17)
\Line(20, 0)(50, 20)
\Line(20,40)(50,20)
\Line(50, 20)(80, 0)
\Line(50, 20)(80, 40)
\end{picture}
\leftrightarrow n_4 &
\begin{picture}(100, 40)(0, 17)
\Line(20, 0)(50, 20)
\Line(20,40)(50,20)
\DashLine(50, 20)(80, 0){3}
\DashLine(50, 20)(80, 40){3}
\end{picture}
\leftrightarrow m_4 &
\begin{picture}(100, 40)(0, 17)
\DashLine(20, 0)(50, 20){3}
\Line(20,40)(50,20)
\Line(50, 20)(80, 0)
\DashLine(50, 20)(80, 40){3}
\end{picture}
\leftrightarrow q_4 \\[10pt]
\begin{picture}(100, 40)(0, 17)
\DashLine(20, 0)(50, 20){3}
\DashLine(20,40)(50,20){3}
\DashLine(50, 20)(80, 0){3}
\DashLine(50, 20)(80, 40){3}
\end{picture}
\leftrightarrow p_4 & &
\begin{picture}(100, 40)(0, 17)
\DashLine(20, 0)(50, 20){3}
\DashLine(20,40)(50,20){3}
\Line(50, 20)(80, 0)
\Line(50, 20)(80, 40)
\end{picture}
\leftrightarrow r_4
\end{array}
\eq\\[5pt]
Here it is understood that the legs pointing to the right are going to be part of the loop. Before we can write down the expression for the 1-loop effective potential, i.e.\ the sum of all 1-loop 1PI diagrams weighed with the appropriate factors, we have to know in how many ways we can connect the internal legs. If we denote the number of vertices that give two solid lines to go into the loop as $p$, the number of vertices that give two dashed lines as $q$ and the number of vertices that give one solid and one dashed line as $r$, then the number of ways to connect these vertices to give a loop is:
\bq
2^{p+q} \frac{\left({r\over2}+p-1\right)!}{\left({r\over2}-1\right)!} \frac{\left({r\over2}+q-1\right)!}{\left({r\over2}-1\right)!} (r-1)! 
\eq

In our case we have of course $p=n_3+n_4+r_4, \; q=m_3+m_4+p_4, \; r=q_3+q_4$. The 1-loop effective potential $V_1$ is now given by:
{\allowdisplaybreaks\bqa
V_1(\eta_1,\eta_2) &=& {1\over(2\pi)^d} \int \; d^dk \sum_{\underset{n_3+n_4+p_4+m_3+m_4+q_3+q_4+r_4\geq1}{n_3,n_4,p_4,m_3,m_4,q_3,q_4,r_4=0}}^{\infty} 3^{n_3} 6^{n_4} 6^{p_4} 2^{q_3} 4^{q_4} \nonumber\\
& & \left(-{3m^2\over\hbar v}\right)^{n_3} \left(-{3m^2\over\hbar v^2}\right)^{n_4} \left(-{3m^2\over\hbar v^2}\right)^{p_4} \left(-{m^2\over\hbar v}\right)^{m_3+q_3} \left(-{m^2\over\hbar v^2}\right)^{m_4+q_4+r_4} \nonumber\\
& & \left({1\over3!}\right)^{n_3} \left({1\over4!}\right)^{n_4} \left({1\over4!}\right)^{p_4} \left({1\over2!}\right)^{m_3+q_3} \left({1\over2!2!}\right)^{m_4+q_4+r_4} \nonumber\\
& & {1\over n_3!} {1\over n_4!} {1\over p_4!} {1\over m_3!} {1\over q_3!} {1\over m_4!} {1\over q_4!} {1\over r_4!} \sum_{n=0}^\infty \delta_{2n,q_3+q_4} \nonumber\\
& & \left(\frac{\hbar}{k^2}\right)^{m_3+m_4+p_4+\h q_3+\h q_4} \left(\frac{\hbar}{k^2+m^2}\right)^{n_3+n_4+r_4+\h q_3+\h q_4} \nonumber\\
& & 2^{n_3+n_4+r_4+m_3+m_4+p_4} \frac{(q_3+q_4-1)!}{(\h q_3+\h q_4-1)!^2} \nonumber\\
& & \left(\h q_3+\h q_4+n_3+n_4+r_4-1\right)! \left(\h q_3+\h q_4+m_3+m_4+p_4-1\right)! \nonumber\\
& & (2p_4+q_3+q_4+2r_4)! (n_3+2n_4+m_3+2m_4+q_4)! \nonumber\\
& & \bigg(-\hbar{1\over(2p_4+q_3+q_4+2r_4)!} {1\over(n_3+2n_4+m_3+2m_4+q_4)!} \cdot \nonumber\\
& & \phantom{\bigg(} \eta_1^{n_3+2n_4+m_3+2m_4+q_4} \eta_2^{2p_4+q_3+q_4+2r_4} \bigg) + \nonumber\\[10pt]
& & -\hbar \;
\begin{picture}(10,10)(0,7)
\Line(0,10)(10,10)
\Vertex(10,10){2}
\end{picture} \;\; \eta_1 - {\hbar\over2} \;
\begin{picture}(20,10)(0,7)
\Line(0,10)(20,10)
\Vertex(10,10){2}
\end{picture} \;\; \eta_1^2 - {\hbar\over2} \;
\begin{picture}(20,10)(0,7)
\DashLine(0,10)(20,10){3}
\Vertex(10,10){2}
\end{picture} \;\; \eta_2^2 - {\hbar\over6} \;
\begin{picture}(20,10)(0,7)
\Line(0,10)(10,10)
\Line(10,10)(20,0)
\Line(10,10)(20,20)
\Vertex(10,10){2}
\end{picture} \;\; \eta_1^3 - {\hbar\over2} \;
\begin{picture}(20,10)(0,7)
\Line(0,10)(10,10)
\DashLine(10,10)(20,0){3}
\DashLine(10,10)(20,20){3}
\Vertex(10,10){2}
\end{picture} \;\; \eta_1\eta_2^2 + \nonumber\\[5pt]
& & -{\hbar\over24} \;
\begin{picture}(20,10)(0,7)
\Line(0,0)(20,20)
\Line(0,20)(20,0)
\Vertex(10,10){2}
\end{picture} \;\; \eta_1^4 - {\hbar\over24} \;
\begin{picture}(20,10)(0,7)
\DashLine(0,0)(20,20){3}
\DashLine(0,20)(20,0){3}
\Vertex(10,10){2}
\end{picture} \;\; \eta_2^4 - {\hbar\over4} \;
\begin{picture}(20,10)(0,7)
\Line(0,0)(20,20)
\DashLine(0,20)(20,0){3}
\Vertex(10,10){2}
\end{picture} \;\; \eta_1^2\eta_2^2
\eqa}

After a long calculation this can be written to:
\bqa
V_1(\f_1,\f_2) &=& {\hbar\over2} {1\over(2\pi)^d} \int \; d^dk \Bigg( \ln\left( 1 + {1\over k^2} \h{m^2\over v^2}(\f_1^2+\f_2^2-v^2) \right) + \nonumber\\
& & \phantom{{\hbar\over2} {1\over(2\pi)^d} \int \; d^dk \Bigg(} \ln\left( 1 + {1\over k^2+m^2} {3\over2}{m^2\over v^2}(\f_1^2+\f_2^2-v^2) \right) \Bigg) + \nonumber\\
& & -\h\left( \delta_\mu|_{\hbar}-{1\over6}v^2 \; \delta_\lambda|_{\hbar} \right) (\f_1^2+\f_2^2-v^2) + {1\over24} \; \delta_\lambda|_{\hbar} \; (\f_1^2+\f_2^2-v^2)^2 \nonumber\\ \label{effpotN2}
\eqa
This result is identical to what Peskin and Schroeder \cite{Peskin} find in their formula (11.74), of course taking into account differences in definitions of coupling constants and counter terms.

\subsubsection{Zero Dimensions}

There is a much quicker, though less straightforward, way to obtain the 1-loop effective potential (\ref{effpotN2}). In zero dimensions it is very easy to find the 1-loop effective action through the Schwinger-Dyson equations. Of course in zero dimensions this effective action is equal to the effective potential. The diagrammatic structure of this 1-loop effective potential in zero dimensions is exactly the same as in $d$ dimensions, only the mathematical expressions corresponding to the diagrams is different. For the 1-loop case however the difference in mathematical expression is not so big: the propagators in the loop, which are $1/m^2$ in zero dimensions just become $1/(k^2+m^2)$ in $d$ dimensions. So if we are able to find the zero-dimensional 1-loop effective potential we can do this replacement to obtain the $d$-dimensional effective potential.

So we first have to calculate the zero-dimensional 1-loop effective potential through the Schwinger-Dyson equations. We write the zero-dimensional action of our $N=2$ LSM generically as:
\bq
S = \h m_1\eta_1^2 + \h m_2\eta_2^2 + {1\over3!}g_1\eta_1^3 + {1\over2!}g_2\eta_1\eta_2^2 + {1\over4!}\lambda_1\eta_1^4 + {1\over4!}\lambda_2\eta_2^4 + {1\over2!2!}\lambda_3\eta_1^2\eta_2^2
\eq
Notice that we have included a mass $\sqrt{m_2}$ for the $\eta_2$-particle now, to be able to do the replacement $m_2\rightarrow k^2$ later. (Also for $m_2=0$ the propagator would not even exist in zero dimensions.)

In diagrammatic form the Schwinger-Dyson equations read:
\bqa
\begin{picture}(20,20)(0,7)
\Line(0,10)(15,10)
\GCirc(15,10){5}{0.5}
\end{picture} &=&
\begin{picture}(20,20)(0,7)
\Line(0,10)(15,10)
\Line(13,8)(17,12)
\Line(13,12)(17,8)
\end{picture} + \h
\begin{picture}(30,20)(0,7)
\Line(0,10)(15,10)
\Line(15,10)(25,20)
\Line(15,10)(25,0)
\GCirc(25,20){5}{0.5}
\GCirc(25,0){5}{0.5}
\end{picture} + \h
\begin{picture}(45,20)(0,7)
\Line(0,10)(15,10)
\CArc(25,10)(10,0,360)
\GCirc(35,10){8}{0.5}
\end{picture} + \h
\begin{picture}(30,20)(0,7)
\Line(0,10)(15,10)
\DashLine(15,10)(25,20){2}
\DashLine(15,10)(25,0){2}
\GCirc(25,20){5}{0.5}
\GCirc(25,0){5}{0.5}
\end{picture} + \h
\begin{picture}(45,20)(0,7)
\Line(0,10)(15,10)
\DashCArc(25,10)(10,0,360){2}
\GCirc(35,10){8}{0.5}
\end{picture} + {1\over3!}
\begin{picture}(40,20)(0,7)
\Line(0,10)(30,10)
\Line(15,10)(25,20)
\Line(15,10)(25,0)
\GCirc(25,20){5}{0.5}
\GCirc(25,0){5}{0.5}
\GCirc(30,10){5}{0.5}
\end{picture} + \nonumber\\
& & \h \begin{picture}(35,20)(0,7)
\Line(0,10)(15,10)
\Line(15,10)(25,20)
\CArc(20,5)(7,0,360)
\GCirc(25,20){5}{0.5}
\GCirc(25,0){5}{0.5}
\end{picture} + {1\over3!}
\begin{picture}(45,20)(0,7)
\Line(0,10)(35,10)
\CArc(25,10)(10,0,360)
\GCirc(35,10){8}{0.5}
\end{picture} + \h
\begin{picture}(40,20)(0,7)
\Line(0,10)(30,10)
\DashLine(15,10)(25,20){2}
\DashLine(15,10)(25,0){2}
\GCirc(25,20){5}{0.5}
\GCirc(25,0){5}{0.5}
\GCirc(30,10){5}{0.5}
\end{picture} + \h
\begin{picture}(35,20)(0,7)
\Line(0,10)(15,10)
\Line(15,10)(25,20)
\DashCArc(20,5)(7,0,360){2}
\GCirc(25,20){5}{0.5}
\GCirc(25,0){5}{0.5}
\end{picture} +
\begin{picture}(35,20)(0,7)
\Line(0,10)(15,10)
\DashLine(15,10)(25,20){2}
\CArc(20,5)(7,0,142)
\DashCArc(20,5)(7,142,360){2}
\GCirc(25,20){5}{0.5}
\GCirc(25,0){5}{0.5}
\end{picture} + \h
\begin{picture}(45,20)(0,7)
\Line(0,10)(35,10)
\DashCArc(25,10)(10,0,360){2}
\GCirc(35,10){8}{0.5}
\end{picture} \nonumber\\[10pt]
\begin{picture}(20,20)(0,7)
\DashLine(0,10)(15,10){2}
\GCirc(15,10){5}{0.5}
\end{picture} &=&
\begin{picture}(20,20)(0,7)
\DashLine(0,10)(15,10){2}
\Line(13,8)(17,12)
\Line(13,12)(17,8)
\end{picture} +
\begin{picture}(30,20)(0,7)
\DashLine(0,10)(15,10){2}
\DashLine(15,10)(25,20){2}
\Line(15,10)(25,0)
\GCirc(25,20){5}{0.5}
\GCirc(25,0){5}{0.5}
\end{picture} +
\begin{picture}(45,20)(0,7)
\DashLine(0,10)(15,10){2}
\CArc(25,10)(10,180,360)
\DashCArc(25,10)(10,0,180){2}
\GCirc(35,10){8}{0.5}
\end{picture} + \h
\begin{picture}(40,20)(0,7)
\DashLine(0,10)(15,10){2}
\DashLine(15,10)(25,20){2}
\Line(15,10)(35,10)
\Line(15,10)(25,0)
\GCirc(25,20){5}{0.5}
\GCirc(25,0){5}{0.5}
\GCirc(30,10){5}{0.5}
\end{picture} + \h
\begin{picture}(35,20)(0,7)
\DashLine(0,10)(15,10){2}
\DashLine(15,10)(25,20){2}
\CArc(20,5)(7,0,360)
\GCirc(25,20){5}{0.5}
\GCirc(25,0){5}{0.5}
\end{picture} + 
\begin{picture}(35,20)(0,7)
\DashLine(0,10)(15,10){2}
\Line(15,10)(25,20)
\CArc(20,5)(7,0,142)
\DashCArc(20,5)(7,142,360){2}
\GCirc(25,20){5}{0.5}
\GCirc(25,0){5}{0.5}
\end{picture} + \nonumber\\
& & \h \begin{picture}(45,20)(0,7)
\DashLine(0,10)(35,10){2}
\CArc(25,10)(10,0,360)
\GCirc(35,10){8}{0.5}
\end{picture} + {1\over3!}
\begin{picture}(40,20)(0,7)
\DashLine(0,10)(30,10){2}
\DashLine(15,10)(25,20){2}
\DashLine(15,10)(25,0){2}
\GCirc(25,20){5}{0.5}
\GCirc(25,0){5}{0.5}
\GCirc(30,10){5}{0.5}
\end{picture} + \h
\begin{picture}(35,20)(0,7)
\DashLine(0,10)(15,10){2}
\DashLine(15,10)(25,20){2}
\DashCArc(20,5)(7,0,360){2}
\GCirc(25,20){5}{0.5}
\GCirc(25,0){5}{0.5}
\end{picture} + {1\over3!}
\begin{picture}(45,20)(0,7)
\DashLine(0,10)(35,10){2}
\DashCArc(25,10)(10,0,360){2}
\GCirc(35,10){8}{0.5}
\end{picture}
\eqa
Here the little crosses indicate the vertices from the sources, respectively $J_1/\hbar$ and $J_2/\hbar$. If we denote the tadpoles by $\phi(J_1,J_2)$ and $\psi(J_1,J_2)$:
\bq
\begin{picture}(20,20)(0,7)
\Line(0,10)(15,10)
\GCirc(15,10){5}{0.5}
\end{picture} \equiv \phi(J_1,J_2) \;, \quad
\begin{picture}(20,20)(0,7)
\DashLine(0,10)(15,10){2}
\GCirc(15,10){5}{0.5}
\end{picture} \equiv \psi(J_1,J_2) \;,
\eq
and their derivatives by
\bqa
\phi_{i_1i_2\ldots i_n} &\equiv& \frac{\partial^n}{\partial J_{i_1} \partial J_{i_2} \ldots \partial J_{i_n}} \; \phi \nonumber\\
\psi_{i_1i_2\ldots i_n} &\equiv& \frac{\partial^n}{\partial J_{i_1} \partial J_{i_2} \ldots \partial J_{i_n}} \; \psi \;,
\eqa
then the Schwinger-Dyson equations read
\bqa
m_1\phi &=& J_1 - \h g_1(\phi^2+\hbar\phi_1) - \h g_2(\psi^2+\hbar\psi_2) - {1\over6}\lambda_1(\phi^3+3\hbar\phi\phi_1+\hbar^2\phi_{11}) + \nonumber\\
& & -\h\lambda_3(\phi\psi^2+\hbar\phi\psi_2+2\hbar\psi\phi_2+\hbar^2\phi_{22}) \nonumber\\
m_2\psi &=& J_2 - g_2(\phi\psi+\hbar\phi_2) - \h\lambda_3(\phi^2\psi+\hbar\phi_1\psi+2\hbar\phi\phi_2+\hbar^2\phi_{12}) + \nonumber\\
& & -{1\over6}\lambda_2(\psi^3+3\hbar\psi\psi_2+\hbar^2\psi_{22})
\eqa

Now the definition of the effective action is
\bqa
\frac{\partial\Gamma}{\partial\phi}(\phi,\psi) &=& J_1(\phi,\psi) \nonumber\\
\frac{\partial\Gamma}{\partial\psi}(\phi,\psi) &=& J_2(\phi,\psi) \;,
\eqa
from which one can derive
\bqa
\phi_1 &=& \frac{-\frac{\partial^2\Gamma}{\partial\psi^2}}{\left(\left(\frac{\partial^2\Gamma}{\partial\phi\partial\psi}\right)^2 - \frac{\partial^2\Gamma}{\partial\phi^2}\frac{\partial^2\Gamma}{\partial\psi^2} \right)} \;, \nonumber\\
\psi_1 &=& \frac{\frac{\partial^2\Gamma}{\partial\phi\partial\psi}}{\left(\left(\frac{\partial^2\Gamma}{\partial\phi\partial\psi}\right)^2 - \frac{\partial^2\Gamma}{\partial\phi^2}\frac{\partial^2\Gamma}{\partial\psi^2} \right)} \;, \nonumber\\
\phi_2 &=& \frac{\frac{\partial^2\Gamma}{\partial\phi\partial\psi}}{\left(\left(\frac{\partial^2\Gamma}{\partial\phi\partial\psi}\right)^2 - \frac{\partial^2\Gamma}{\partial\phi^2}\frac{\partial^2\Gamma}{\partial\psi^2} \right)} \;, \nonumber\\
\psi_2 &=& \frac{-\frac{\partial^2\Gamma}{\partial\phi^2}}{\left(\left(\frac{\partial^2\Gamma}{\partial\phi\partial\psi}\right)^2 - \frac{\partial^2\Gamma}{\partial\phi^2}\frac{\partial^2\Gamma}{\partial\psi^2} \right)} \;.
\eqa
Through these relations we can write the Schwinger-Dyson equations in terms of (partial derivatives of) the effective action and the tadpole. Then it appears one can solve these partial differential equations iteratively up to some order to express the effective action in terms of the tadpole. Assuming that the effective action starts with a term of order $\hbar^0$, which is characteristic of the canonical approach, and writing
\bq
\Gamma(\phi,\psi) = A(\phi,\psi) + \hbar B(\phi,\psi) + \ldots\;
\eq
we find:
\bqa
A(\phi,\psi) &=& \h m_1\phi^2 + \h m_2\psi^2 + {1\over6}g_1\phi^3 + {1\over2}g_2\phi\psi^2 + {1\over24}\lambda_1\phi^4 + {1\over24}\lambda_2\psi^4 + {1\over4}\lambda_3\phi^2\psi^2 \nonumber\\
B(\phi,\psi) &=& \h \ln\left( \frac{\partial^2A}{\partial\phi^2} \frac{\partial^2A}{\partial\psi^2} - \left(\frac{\partial^2A}{\partial\phi\partial\psi}\right)^2 \right) + C \label{effactAB}
\eqa
Here the constant $C$ is just a constant of integration, which is unimportant for the physics. It is convenient to fix it however by demanding that for a free theory $B=0$, which gives $C=-\h\ln m_1m_2$.

Now to obtain the 1-loop effective potential in $d$ dimensions (excluding counter terms) we have to make the replacements
\bq
m_1 \rightarrow k^2+m^2 \;, \quad m_2 \rightarrow k^2 \;,
\eq
in $\hbar B(\phi,\psi)$ and add the integration ${1\over(2\pi)^d}\int d^dk$. 

If we do this, specify all the masses and coupling constants $m$, $g$ and $\lambda$ to the masses and coupling constants we have in the $N=2$ LSM, and write the $\eta_1$- and $\eta_2$-field in terms of the $\f_1$- and $\f_2$-field again we find exactly (\ref{effpotN2}), of course excluding the counter terms.

Also notice that (\ref{effactAB}) shows in general (for a 2-field theory) that the effective potential becomes complex when the classical potential $A$ becomes non-convex. Inside the logarithm in $B$ in (\ref{effactAB}) is the Hessian of the function $A$, which is negative where the function $A$ is non-convex.

\subsubsection{Calculating The Effective Potential}

To proceed calculating (\ref{effpotN2}) we have to expand the logarithms again to let any divergent parts cancel the divergences in the counter terms. Of course when we expand the logarithm with the $1/k^2$ a lot of infrared divergences are going to appear. These divergences should later sum up to something finite again, but for the moment we have to regularize them, which we do by introducing a mass $\varepsilon$ for the $\eta_2$-particle. The 1-loop effective potential becomes:
\bqa
V_1 &=& \bigg( {1\over16}{\hbar m^4\over v^4} \left( I(0,m,0,m) - I(0,\varepsilon,0,\varepsilon) \right) - 7{\hbar m^6\over v^4} \; I(0,m,0,m,0,m) + \nonumber\\
& & \phantom{\bigg(} {41\over4}{\hbar m^8\over v^4} \; I(0,m,0,m,0,m,0,m) \bigg) (\f_1^2+\f_2^2-v^2)^2 + \nonumber\\
& & -{\hbar\over2} \sum_{n=3}^\infty {1\over n} \left( -\h{m^2\over v^2}(\f_1^2+\f_2^2-v^2) \right)^n {1\over(2\pi)^d} \int d^dk \; \frac{1}{(k^2+\varepsilon^2)^n} + \nonumber\\
& & -{\hbar\over2} \sum_{n=3}^\infty {1\over n} \left( -{3\over2}{m^2\over v^2}(\f_1^2+\f_2^2-v^2) \right)^n {1\over(2\pi)^d} \int d^dk \; \frac{1}{(k^2+m^2)^n}
\eqa
In the first term the ultraviolet divergences cancel, we can write this term as:
\bqa
& & I(0,m,0,m) - I(0,\varepsilon,0,\varepsilon) = \nonumber\\
& & \qquad 2(\varepsilon^2-m^2) \; I(0,m,0,m,0,\varepsilon) - (\varepsilon^2-m^2)^2 \; I(0,m,0,m,0,\varepsilon,0,\varepsilon)
\eqa

Now using that for $d\leq4$ and $n\geq3$ we have
\bq
{1\over(2\pi)^d} \int d^dk \; {1\over\left(k^2+m^2\right)^n} = {1\over(4\pi)^{d/2}} m^{d-2n} {\Gamma(n-d/2)\over\Gamma(n)} \;,
\eq
and for $d\leq4$ and $n+p\geq3$ we have
\bqa
& & {1\over(2\pi)^d} \int d^dk \; {1\over\left(k^2+m^2\right)^n} {1\over\left(k^2+\varepsilon^2\right)^p} = \nonumber\\
& & \qquad {1\over(4\pi)^{d/2}} \frac{\Gamma(n+p-d/2)}{\Gamma(n)\Gamma(p)} \int_0^1 dx \; x^{n-1} (1-x)^{p-1} (xm^2+(1-x)\varepsilon^2)^{d/2-n-p} \;, \nonumber\\
\eqa
we find for the 1-loop effective potential
\bqa
V_1 &=& \bigg( {1\over16}{\hbar m^4\over v^4} \left(m^{d-4}-\varepsilon^{d-4}\right) {1\over(4\pi)^{d/2}} \Gamma(2-d/2) - {7\over2}{\hbar m^d\over v^4} {1\over(4\pi)^{d/2}} \Gamma(3-d/2) + \nonumber\\
& & \phantom{\bigg(} {41\over24}{\hbar m^d\over v^4} {1\over(4\pi)^{d/2}} \Gamma(4-d/2) \bigg) (\f_1^2+\f_2^2-v^2)^2 + \nonumber\\
& & -{1\over(4\pi)^{d/2}} \varepsilon^d {\hbar\over2} \sum_{n=3}^\infty {1\over n!} \left( -\h{m^2\over v^2\varepsilon^2}(\f_1^2+\f_2^2-v^2) \right)^n \Gamma(n-d/2) + \nonumber\\
& & -{1\over(4\pi)^{d/2}} m^d {\hbar\over2} \sum_{n=3}^\infty {1\over n!} \left( -{3\over2}{1\over v^2}(\f_1^2+\f_2^2-v^2) \right)^n \Gamma(n-d/2) \;. \label{effpotN2ddim}
\eqa

We see that for $d\leq4$ all ultraviolet divergences cancel, which shows again that the theory is renormalizable for $d\leq4$.

To find a more explicit expression for $V_1$ we have to specify the dimension $d$.

\subsubsection{$d=1$ And $d=2$}

If one substitutes $d=1$ in (\ref{effpotN2ddim}), performs the sums and works everything out one finds that the divergences for $\varepsilon\rightarrow0$ do \emph{not} cancel. The same happens for $d=2$. This is generally known, in one and two dimensions there is \emph{no} SSB, which is manifested by the remaining infrared divergences. See for example Coleman \cite{Coleman} and Coleman, Jackiw and Politzer \cite{Jackiw}.

\subsubsection{$d=4$}

\begin{figure}[h]
\begin{center}
\epsfig{file=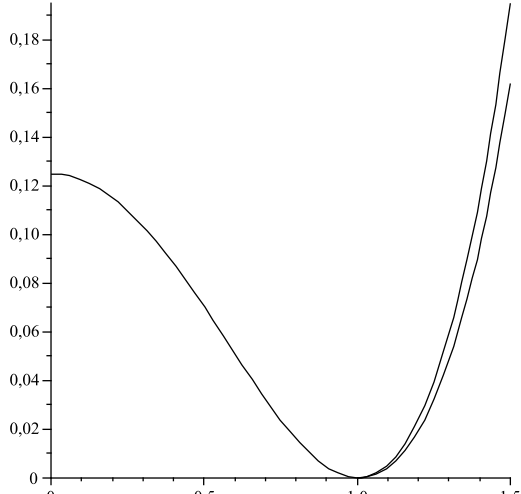,width=7cm}
\end{center}
\vspace{-1cm}
\caption{$V_0$ And $V=V_0+V_1$ as a function of $\sqrt{\f_1^2+\f_2^2}/v$ for $\hbar=2$, $m=1$, $v=1$.}
\label{effpotN21mind4}
\end{figure}

In $d=4$ the infrared divergences do cancel and one finds:
\bqa
V_1 &=& -{3\over8}{\hbar m^4\over v^2}{1\over16\pi^2} (\f_1^2+\f_2^2-v^2) - {131\over48}{\hbar m^4\over v^4}{1\over16\pi^2} (\f_1^2+\f_2^2-v^2)^2 + \nonumber\\
& & {1\over16}{\hbar m^4\over v^4}{1\over16\pi^2} (\f_1^2+\f_2^2-v^2)^2 \ln\left( \h{1\over v^2}(\f_1^2+\f_2^2-v^2) \right) + \nonumber\\
& & {1\over4}\hbar m^4{1\over16\pi^2} \left( 1+{3\over2}{1\over v^2}(\f_1^2+\f_2^2-v^2) \right)^2 \ln\left( 1+{3\over2}{1\over v^2}(\f_1^2+\f_2^2-v^2) \right) \label{effpotN2d4}
\eqa
In figure \ref{effpotN21mind4} the complete effective potential (up to one loop) and the classical potential are plotted for the case $\hbar=2$, $m=1$, $v=1$.

The minimum of the effective potential is at $\f_1^2+\f_2^2=v^2$, as our tadpole renormalization condition in (\ref{rencond2}) ensures. Also exactly at this point the effective potential becomes complex because of the first logarithm in (\ref{effpotN2d4}). This shows again that the effective potential in the canonical approach becomes complex where the classical potential becomes non-convex. Because the effective potential becomes complex exactly at the location of the minima we \emph{cannot} compute the $n$-points Green's functions from it. This is related to the fact that all these $n$-points Green's functions suffer from infrared divergences at zero incoming momentum.

Note that the effective potential we have computed here is \emph{convex} where it is defined. This is \emph{not} always the case. We could easily have chosen other renormalization conditions such that the minima of the effective potential occur for $\f_1^2+\f_2^2>v^2$ (by for example adding a constant term to $\delta_\mu|_{\hbar}$). Then there is a non-convex region between these minima and the circle $\f_1^2+\f_2^2=v^2$, where the effective potential becomes complex again. In fact in Peskin and Schroeder \cite{Peskin} such an effective potential is found in (11.79). They use the $\overline{MS}$ renormalization scheme. Their remark that fortunately the minima of the effective potential occur outside the region where it becomes complex is somewhat inappropriate, since we have shown here that this is \emph{not} always the case.

So for the $N=2$ LSM there is an apparent convexity problem. Again, as has been thoroughly discussed in the literature this problem is resolved by realizing that the canonical and path-integral approach are \emph{not} the same in the case of a non-convex classical action.

\section{The Path-Integral Approach I}\label{pathI}

In this section we will discuss the path-integral approach to the Euclidean $N=2$ LSM. This means we want to calculate the path integral of this model in some approximation. For models for which all minima are nicely seperated the path integral can be calculated with a saddle-point approximation. See for example \cite{Fujimoto,Bender,Cooper} and chapter 4 of \cite{vanKessel}. This means we expand the generating functional around each minimum and add all these generating functionals to obtain the complete generating functional. In the examples just mentioned this is a good approximation because the minima lie far away from each other. In the $N=2$-case we can, in principle, also use such a saddle-point approximation, however now the minima form a continuous set and do \emph{not} lie far apart. So it is questionable whether expanding around each minimum and then summing, or rather integrating, the contributions from each minimum gives a reasonable approximation to the path integral.

Another questionable point is the perturbative expansion around each minimum. When making this expansion one has replaced the, in principle damped, $\eta_2$-direction (i.e.\ tangential direction) by a non-damped straight line. There is an $\eta_2^4$-term that damps oscillations in the $\eta_2$-direction in principle, however in perturbation theory the exponential of this term is expanded, and \emph{not} all terms are kept. In this way we lose the damping effect in the tangential direction, which \emph{is} actually there.

In this section we shall just perform the naive saddle-point approximation, even though the arguments above advise strongly against it. There is also an argument in favor of this naive approach. We know that expanding around \emph{one} minimum (i.e.\ the canonical approach) gives a self-consistent theory and the Green's functions calculated in this way satisfy the Schwinger-Dyson equations. Also the generating functional calculated by including only \emph{one} minimum satisfies the Schwinger-Symanzik equations. Because the Schwinger-Dyson and Schwinger-Symanzik equations are linear (in the \emph{full} Green's functions or generating functional) also the sum of several \emph{full} Green's functions or generating functionals around different minima are solutions to these equations. So we know at least that the full Green's functions and generating functional obtained by summing or integrating over \emph{all} minima are solutions to the Schwinger-Dyson and Schwinger-Symanzik equations. 

\subsection{Green's Functions}

The renormalized action of the $d$-dimensional $N=2$ LSM is:
\bqa
S &=& \int d^dx \; \bigg( \h\left(\nabla\f_1\right)^2 + \h\left(\nabla\f_2\right)^2 - \h\mu\left(\f_1^2+\f_2^2\right) + {\lambda\over24}\left(\f_1^2+\f_2^2\right)^2 + \nonumber\\
& & \phantom{\int d^dx \; \bigg(} \h\delta_Z\left(\nabla\f_1\right)^2 + \h\delta_Z\left(\nabla\f_2\right)^2 - \h\delta_{\mu}\left(\f_1^2+\f_2^2\right) + {\delta_{\lambda}\over24}\left(\f_1^2+\f_2^2\right)^2 \bigg) \;.
\eqa
The minima of the first line are given by
\bqa
\f_1 &=& v\cos\delta \equiv v_1(\delta) \nonumber\\
\f_2 &=& v\sin\delta \equiv v_2(\delta) \;,
\eqa
with $v=\sqrt{6\mu/\lambda}$ again. Now we expand the action around \emph{one} of these minima:
\bq
\f_1 = v_1 + \eta_1 \;, \quad \f_2 = v_2 + \eta_2
\eq
When writing the action in terms of these $\eta$-fields the Gaussian part becomes non-diagonal in $\eta_1$ and $\eta_2$. To make this part diagonal again we introduce the $\psi$-fields:
\bqa
\psi_1 &=& {1\over v}(v_1\eta_1+v_2\eta_2) \nonumber\\
\psi_2 &=& {1\over v}(v_2\eta_1-v_1\eta_2)
\eqa
\bqa
\eta_1 &=& {1\over v}(v_1\psi_1+v_2\psi_2) \nonumber\\
\eta_2 &=& {1\over v}(v_2\psi_1-v_1\psi_2)
\eqa
In terms of these $\psi$-fields the action reads (again defining $\mu\equiv\h m^2$):
\bqa
S &=& \int d^dx \; \bigg( \h\left(\nabla\psi_1\right)^2 + \h\left(\nabla\psi_2\right)^2 + \h m^2\psi_1^2 + \nonumber\\
& & \phantom{\int d^dx \; \bigg(} \h{m^2\over v}\psi_1^3 + \h{m^2\over v}\psi_1\psi_2^2 + {1\over8}{m^2\over v^2}\psi_1^4 + {1\over8}{m^2\over v^2}\psi_2^4 + {1\over4}{m^2\over v^2}\psi_1^2\psi_2^2 + \nonumber\\
& & \phantom{\int d^dx \; \bigg(} \h\delta_Z\left(\nabla\psi_1\right)^2 + \h\delta_Z\left(\nabla\psi_2\right)^2 + \left(-v\delta_\mu+{1\over6}v^3\delta_\lambda\right)\psi_1 + \nonumber\\
& & \phantom{\int d^dx \; \bigg(} \left(-\h\delta_\mu+{1\over4}v^2\delta_\lambda\right)\psi_1^2 + \left(-\h\delta_\mu+{1\over12}v^2\delta_\lambda\right)\psi_2^2 + \nonumber\\
& & \phantom{\int d^dx \; \bigg(} {1\over6}v\delta_\lambda\psi_1^3 + {1\over6}v\delta_\lambda\psi_1\psi_2^2 + {\delta_\lambda\over24}\psi_1^4 + {\delta_\lambda\over24}\psi_2^4 + {\delta_\lambda\over12}\psi_1^2\psi_2^2 \bigg)
\eqa

Notice that this action does \emph{not} depend on $\delta$ anymore, as is expected from the $O(2)$-invariance of this model. Also notice that the action for the $\psi$-fields is exactly the same as the action for the $\eta$-fields in the canonical approach (\ref{N2etaaction}). This means the $\psi$-Green's functions are also identical to the $\eta$-Green's functions in the canonical approach, and for these Green's functions we can use the results from the previous section.

Now we wish to obtain the $\f$-Green's functions. As stated in the introduction we are going to calculate these by just integrating over the contributions from all minima, i.e.\ integrate over $\delta$. One should keep in mind here that the $\psi$-Green's functions do \emph{not} depend on $\delta$ anymore.
\bqa
\langle \f_1(x) \rangle &=& {1\over2\pi} \int_{-\pi}^\pi d\delta \; \left( v_1 + \langle \eta_1(x) \rangle \right) = {1\over2\pi} \int_{-\pi}^\pi d\delta \; \left( v_1 + {v_1\over v}\langle \psi_1(x) \rangle + {v_2\over v} \langle \psi_2(x) \rangle \right) = 0 \nonumber\\
\langle \f_2(x) \rangle &=& {1\over2\pi} \int_{-\pi}^\pi d\delta \; \left( v_2 + \langle \eta_2(x) \rangle \right) = {1\over2\pi} \int_{-\pi}^\pi d\delta \; \left( v_2 + {v_2\over v}\langle \psi_1(x) \rangle - {v_1\over v} \langle \psi_2(x) \rangle \right) = 0 \nonumber\\
\eqa
\bqa
\langle \f_1(x)\f_1(y) \rangle &=& {1\over2\pi} \int_{-\pi}^\pi d\delta \; \langle \left(v_1 + \eta_1(x)\right)\left(v_1 + \eta_1(y)\right) \rangle \nonumber\\
&=& \h v^2 + \h v \langle\psi_1(x)\rangle + \h v \langle\psi_1(y)\rangle + \h \langle\psi_1(x)\psi_1(y)\rangle + \h \langle\psi_2(x)\psi_2(y)\rangle \nonumber\\
\langle \f_2(x)\f_2(y) \rangle &=& \h v^2 + \h v \langle\psi_1(x)\rangle + \h v \langle\psi_1(y)\rangle + \h \langle\psi_1(x)\psi_1(y)\rangle + \h \langle\psi_2(x)\psi_2(y)\rangle \nonumber\\
\langle \f_1(x)\f_2(y) \rangle &=& 0 \label{Cartavintermspsi}
\eqa
In this last line we also used $\langle\psi_2(x)\rangle = \langle\psi_1(x)\psi_2(y)\rangle = 0$.

With the results of the previous chapter it is now easy to obtain the $\f_1$- and $\f_2$-propagator up to 1-loop order. If we use the same counter terms as in the canonical approach, we have, up to 1-loop order (using the tadpole renormalization condition (\ref{rencond2}) and the Dyson summed propagators (\ref{eta1prop}) and (\ref{eta2prop})):
\bqa
\langle\psi_1(x)\rangle &=& 0 \nonumber\\
\langle\psi_1(x)\psi_1(y)\rangle &=& {1\over(2\pi)^d} \int d^dp \; e^{ip\cdot(x-y)} \; \frac{\hbar}{p^2+m^2-\hbar A_1(p^2)} \nonumber\\
\langle\psi_2(x)\psi_2(y)\rangle &=& {1\over(2\pi)^d} \int d^dp \; e^{ip\cdot(x-y)} \; \frac{\hbar}{p^2-\hbar A_2(p^2)}
\eqa
with $A_1$ and $A_2$ given in (\ref{A1canon}) and (\ref{A2canon}). Finally we find:
\bqa
\langle \f_1(x)\f_1(y) \rangle = \langle \f_2(x)\f_2(y) \rangle &=& \h v^2 + \nonumber\\
& & \h {1\over(2\pi)^d} \int d^dp \; e^{ip\cdot(x-y)} \; \frac{\hbar}{p^2+m^2-\hbar A_1(p^2)} + \nonumber\\
& & \h {1\over(2\pi)^d} \int d^dp \; e^{ip\cdot(x-y)} \; \frac{\hbar}{p^2-\hbar A_2(p^2)} \;.
\eqa

With the formulas (\ref{Cartavintermspsi}) it is also easy to calculate the $\f_1$- and $\f_2$-propagator up to order $\hbar^2$. All we need more is $\langle\psi_1(x)\rangle$ at order $\hbar^2$. This quantity can easily be calculated with the Feynam rules from section \ref{can}. In this case we shall not specify the counter terms, but keep them general. This will later be convenient when comparing the upcoming result for the $\f_1$- and $\f_2$-propagator to the result obtained from a calculation via the path integral in terms of polar field variables. $\langle\psi_1(x)\rangle$ At order $\hbar^2$ is now:
{\allowdisplaybreaks\bqa
\langle \tilde{\psi}_1 \rangle|_{\hbar^2} &=& \quad
\begin{picture}(70,40)(0,18)
\Line(0,20)(20,20)
\GCirc(55,20){15}{0.7}
\DashCArc(37,20)(17,50,180){3}
\DashCArc(37,20)(17,180,310){3}
\end{picture} \quad + \quad
\begin{picture}(70,40)(0,18)
\Line(0,20)(20,20)
\GCirc(55,20){15}{0.7}
\CArc(37,20)(17,50,180)
\CArc(37,20)(17,180,310)
\end{picture} \quad + \quad
\quad \begin{picture}(70,40)(0,18)
\Line(0,20)(40,20)
\GCirc(55,20){15}{0.7}
\CArc(20,30)(10,-90,270)
\end{picture} \quad + \nonumber\\[10pt]
& & \quad \begin{picture}(70,40)(0,18)
\Line(0,20)(40,20)
\GCirc(55,20){15}{0.7}
\DashCArc(20,30)(10,-90,270){3}
\end{picture} \quad + \quad
\quad \begin{picture}(70,40)(0,18)
\Line(0,20)(25,20)
\Line(25,20)(60,33)
\Line(25,20)(60,7)
\GCirc(60,33){10}{0.7}
\GCirc(60,7){10}{0.7}
\end{picture} \quad + \quad
\begin{picture}(70,40)(0,18)
\Line(0,20)(70,20)
\DashCArc(50,20)(20,-180,180){3}
\end{picture} \quad + \nonumber\\[10pt]
& & \quad \begin{picture}(70,40)(0,18)
\Line(0,20)(70,20)
\CArc(50,20)(20,-180,180)
\end{picture} \quad + \quad
\begin{picture}(70,40)(0,18)
\Line(0,20)(50,20)
\Vertex(50,20){3}
\end{picture} \quad + \quad
\begin{picture}(70,40)(0,18)
\Line(0,20)(40,20)
\GCirc(55,20){15}{0.7}
\Vertex(20,20){3}
\end{picture} \quad + \nonumber\\[10pt]
& & \quad \begin{picture}(70,40)(0,18)
\Line(0,20)(40,20)
\CArc(55,20)(15,0,360)
\Vertex(40,20){3}
\end{picture} \quad + \quad
\begin{picture}(70,40)(0,18)
\Line(0,20)(40,20)
\DashCArc(55,20)(15,-180,180){3}
\Vertex(40,20){3}
\end{picture} \nonumber\\[30pt]
&=& -{1\over8} {\hbar^2\over v^3} \; I(0,0)^2 - {3\over4} {\hbar^2\over v^3} \; I(0,0) I(0,m) - {9\over8} {\hbar^2\over v^3} \; I(0,m)^2 \nonumber\\
& & -{3\over2} {\hbar^2m^2\over v^3} \; I(0,0) I(0,m,0,m) - {9\over2} {\hbar^2m^2\over v^3} \; I(0,m) I(0,m,0,m) + \nonumber\\
& & -\h {\hbar^2m^2\over v^3} \; I(0,m) I(0,0,0,0) + \h {\hbar^2m^2\over v^3} \; I(0,0) I(0,0,0,0) \nonumber\\
& & +\h{\hbar^2m^2\over v^3} \; D_{00m} + {3\over2} {\hbar^2m^2\over v^3} \; D_{mmm} - {3\over4} {\hbar^2m^4\over v^3} \; B_{m00} - {27\over4} {\hbar^2m^4\over v^3} \; B_{mmm} \nonumber\\
& & -\h {\hbar^2m^4\over v^3} \; B_{00m} \nonumber\\
& & +{3\over2} {\hbar\over v} \; I(0,m) \; \delta_Z|_{\hbar} + {3\over2} {\hbar\over m^2v} \; I(0,m) \; \delta_\mu|_{\hbar} - {1\over4} {\hbar v\over m^2} \; I(0,m) \; \delta_{\lambda}|_{\hbar} \nonumber\\
& & +\h {\hbar\over v} \; I(0,0) \; \delta_Z|_{\hbar} + \h {\hbar\over m^2v} \; I(0,0) \; \delta_\mu|_{\hbar} - {1\over12} {\hbar v\over m^2} \; I(0,0) \; \delta_{\lambda}|_{\hbar} \nonumber\\
& & -{3\over2} {\hbar m^2\over v} \; I(0,m,0,m) \; \delta_Z|_{\hbar} + 3 {\hbar\over v} \; I(0,m,0,m) \; \delta_\mu|_{\hbar} + \nonumber\\
& & -\h {v\over m^4} \; \left(\delta_\mu|_{\hbar}\right)^2 + {1\over24} {v^5\over m^4} \; \left(\delta_{\lambda}|_{\hbar}\right)^2 - {1\over6} {v^3\over m^4} \; \delta_\mu|_{\hbar} \delta_{\lambda}|_{\hbar} + {v\over m^2} \; \delta_\mu|_{\hbar^2} - {1\over6} {v^3\over m^2} \; \delta_{\lambda}|_{\hbar^2} \nonumber\\
\eqa}

Here we have expressed everything in terms of the standard integrals listed in appendix \ref{appstandint}.

Substituting this and the already obtained $\psi_1$- and $\psi_2$-propagator in (\ref{Cartavintermspsi}) gives:
{\allowdisplaybreaks\bqa
\langle \varphi_1(0) \varphi_1(x) \rangle_{\mathrm{c}} &=& +\h v^2 \nonumber\\
& & -\h \hbar \; I(0,0) + \h \hbar \; A_0(x) - {3\over2} \hbar \; I(0,m) + \h \hbar \; A_m(x) \nonumber\\
& & +{v^2\over m^2} \; \delta_\mu|_{\hbar} - {1\over6} {v^4\over m^2} \; \delta_{\lambda}|_{\hbar} \nonumber\\
& & +\h {\hbar^2m^2\over v^2} \; I(0,0) I(0,0,0,0) - \h {\hbar^2m^2\over v^2} \; I(0,m) I(0,0,0,0) \nonumber\\
& & -{3\over2} {\hbar^2m^2\over v^2} \; I(0,0) I(0,m,0,m) - {9\over2} {\hbar^2m^2\over v^2} \; I(0,m) I(0,m,0,m) \nonumber\\
& & +\h {\hbar^2m^2\over v^2} \; D_{00m} + {3\over2} {\hbar^2m^2\over v^2} \; D_{mmm} - \h {\hbar^2m^4\over v^2} \; B_{00m} - {3\over4} {\hbar^2m^4\over v^2} \; B_{m00} \nonumber\\
& & -{27\over4} {\hbar^2m^4\over v^2} \; B_{mmm} \nonumber\\
& & -\h {\hbar^2m^2\over v^2} \; I(0,0) C_{00}(x) + \h {\hbar^2m^2\over v^2} \; I(0,m) C_{00}(x) \nonumber\\
& & +\h {\hbar^2m^2\over v^2} \; I(0,0) C_{mm}(x) + {3\over2} {\hbar^2m^2\over v^2} \; I(0,m) C_{mm}(x) \nonumber\\
& & +\h {\hbar^2m^4\over v^2} \; B_{00m}(x) + {1\over4} {\hbar^2m^4\over v^2} \; B_{m00}(x) + {9\over4} {\hbar^2m^4\over v^2} \; B_{mmm}(x) \nonumber\\
& & +\h \hbar \; I(0,0) \; \delta_Z|_{\hbar} - \h \hbar \; A_0(x) \; \delta_Z|_{\hbar} + {3\over2} \hbar \; I(0,m) \; \delta_Z|_{\hbar} - \h \hbar \; A_m(x) \; \delta_Z|_{\hbar} \nonumber\\
& & -{3\over2} \hbar m^2 \; I(0,m,0,m) \; \delta_Z|_{\hbar} + 3 \hbar \; I(0,m,0,m) \; \delta_\mu|_{\hbar} \nonumber\\
& & +\h \hbar m^2 \; C_{mm}(x) \; \delta_Z|_{\hbar} - \hbar \; C_{mm}(x) \; \delta_\mu|_{\hbar} \nonumber\\
& & +{1\over18}{v^6\over m^4} \; \left(\delta_{\lambda}|_{\hbar}\right)^2 - {1\over3}{v^4\over m^4} \; \delta_\mu|_{\hbar} \; \delta_{\lambda}|_{\hbar} + {v^2\over m^2} \; \delta_\mu|_{\hbar^2} - {1\over6}{v^4\over m^2} \; \delta_{\lambda}|_{\hbar^2} \nonumber\\[5pt]
& & +\mathcal{O}\left(\hbar^3\right) \label{phi1propchap7}
\eqa}

\subsection{The Effective Potential}

Now we will try to find the effective potential of the $N=2$ LSM. To this end we introduce source terms in the action. Because we are only interested in the effective \emph{potential} we shall take the sources to be constant over space time. Including these source terms the action is:
\bqa
S &=& \int d^dx \; \bigg( \h\left(\nabla\f_1\right)^2 + \h\left(\nabla\f_2\right)^2 - \h\mu\left(\f_1^2+\f_2^2\right) + {\lambda\over24}\left(\f_1^2+\f_2^2\right)^2 - J_1\f_1 - J_2\f_2 + \nonumber\\
& & \phantom{\int d^dx \; \bigg(} \h\delta_Z\left(\nabla\f_1\right)^2 + \h\delta_Z\left(\nabla\f_2\right)^2 - \h\delta_{\mu}\left(\f_1^2+\f_2^2\right) + {\delta_{\lambda}\over24}\left(\f_1^2+\f_2^2\right)^2 \bigg) \;.
\eqa

Now we have to find the minima of the first line again. Only for the case $J_1=J_2=0$ we have a ring of minima, as found in the previous section. For one of the sources non-zero however there is only \emph{one} minimum (and \emph{one} saddle point). This means that when both sources are of order $\hbar^0$ taking into account \emph{one} minimum is a good approximation. Below we shall show that taking into account \emph{one} minimum is equivalent to the canonical approach, outlined in the previous section. However, when the sources become of order $\hbar$ the minimum becomes so unstable that quantum fluctuations along the ring become important. Clearly in this regime it is a bad approximation to take into account only this single minimum, although it is the \emph{only} true minimum (for $J\neq0$). In this regime we have to take notice of all the points in the ring. What all the points in the ring have in common is that they are minima in $r=\sqrt{\f_1^2+\f_2^2}$. So to find these points, also for non-zero sources, we have to minimize the classical action with respect to $r$. Writing the classical field as:
\bqa
\f_1 &=& r\cos\delta \nonumber\\
\f_2 &=& r\sin\delta \;, \label{classpoints}
\eqa
we find the equation
\bq
-\mu r + {\lambda\over6}r^3 - J_1\cos\delta - J_2\sin\delta = 0 \;.
\eq
Writing
\bqa
J_1 &=& J\cos\beta \nonumber\\
J_2 &=& J\sin\beta
\eqa
and parameterizing $J$ as
\bq
J = {2\mu v\over3\sqrt{3}} \frac{\sin(3\alpha)}{\cos(\delta-\beta)} \;, \quad \alpha = 0,\ldots,{\pi\over6}
\eq
we find the solution
\bq \label{r}
r = {2v\over\sqrt{3}} \sin\left(\alpha+{\pi\over3}\right) \;.
\eq
So for each angle $\delta$ we have a point on the ring given by (\ref{r}).

Again we should expand the action around the classical points (\ref{classpoints}). To make the action diagonal we have to introduce the $\psi$-fields again:
\bqa
\psi_1 &=& \cos\delta \; \eta_1 + \sin\delta \; \eta_2 \nonumber\\
\psi_2 &=& \sin\delta \; \eta_1 - \cos\delta \; \eta_2
\eqa
The action becomes:
{\allowdisplaybreaks\bqa
S &=& \int d^dx \; \bigg( \h\mu r^2 - {3\over4}{\mu\over v^2}r^4 - \h\delta_\mu r^2 + {\delta_\lambda\over24}r^4 + \nonumber\\
& & \phantom{\int d^dx \; \bigg(} \h\left(\nabla\psi_1\right)^2 + \h\left(\nabla\psi_2\right)^2 + \left( -\h\mu+{3\over2}{\mu\over v^2}r^2 \right)\psi_1^2 \nonumber\\
& & \phantom{\int d^dx \; \bigg(} -J\sin(\delta-\beta)\psi_2 + \left( -\h\mu+\h{\mu\over v^2}r^2 \right)\psi_2^2 + \nonumber\\
& & \phantom{\int d^dx \; \bigg(} {\mu\over v^2}r\psi_1^3 + {\mu\over v^2}r\psi_1\psi_2^2 + {1\over4}{\mu\over v^2}\psi_1^4 + {1\over4}{\mu\over v^2}\psi_2^4 + \h{\mu\over v^2}\psi_1^2\psi_2^2 + \nonumber\\
& & \phantom{\int d^dx \; \bigg(} \h\delta_Z\left(\nabla\psi_1\right)^2 + \h\delta_Z\left(\nabla\psi_2\right)^2 + \left(-\delta_\mu r+{1\over6}\delta_\lambda r^3\right) \psi_1 + \nonumber\\
& & \phantom{\int d^dx \; \bigg(} \left(-\h\delta_\mu+{1\over4}\delta_\lambda r^2\right) \psi_1^2 + \left(-\h\delta_\mu+{1\over12}\delta_\lambda r^2\right) \psi_2^2 \nonumber\\
& & \phantom{\int d^dx \; \bigg(} {1\over6}\delta_\lambda r \psi_1^3 + {1\over6}\delta_\lambda r \psi_1\psi_2^2 + {1\over24}\delta_\lambda \psi_1^4 + {1\over24}\delta_\lambda \psi_2^4 + {1\over12}\delta_\lambda \psi_1^2\psi_2^2 \bigg)
\eqa}

Now we shall take the magnitude of the source $J=\sqrt{J_1^2+J_2^2}$ to be of order $\hbar$. To proceed further with the calculation one has to make an approximation. The most straightforward option is to treat all terms of order higher than $\hbar$ in the action as a perturbation. This means we should also expand $r$ in $\hbar$:
\bqa
r(\delta) &=& {2v\over\sqrt{3}} \sin\left( {1\over3}\arcsin\left({3\sqrt{3}\over2\mu v}J\cos(\delta-\beta)\right) + {\pi\over3} \right) \nonumber\\
&=& v + {1\over2\mu}J\cos(\delta-\beta) - {3\over8}{1\over v\mu^2}J^2\cos^2(\delta-\beta) + \mathcal{O}(\hbar^3) \label{r2}
\eqa
Then one can read off the Feynman rules from the action and calculate the generating functional and the $\psi_1$- and $\psi_2$-tadpole with Feynman diagrams. This is all straightforward, but at the end one finds an infrared-divergent expression. One might have expected this from the results of the previous chapter. There we saw that, in $d=3$ and $d=4$, the infrared divergences only sum up to something finite if we include \emph{all} 1-loop graphs. Because we take $J$ of order $\hbar$ here it means effectively that we \emph{cannot} calculate \emph{any} $n$-points Green's functions from our generating functional. For this one would need to know the exact $J$-dependence. This in turn means we are \emph{not} including all 1-loop graphs and we cannot expect the infrared divergences to disappear.

Another thing one can do, which is less straightforward, but gives results without remaining infrared divergences, is ignore the term
\bq \label{discpsi2term}
-J\sin(\delta-\beta)\psi_2
\eq
in the action, because $J$ is small anyway. Then $\psi_1^2$ and $\psi_2^2$ are of order $\hbar$ and we shall only keep the Gaussian terms. Doing this we find for the generating functional:
\bqa
Z_{\delta} &=& \exp\left(-{1\over\hbar}\Omega\left( \h\mu r^2 - {3\over4}{\mu\over v^2}r^4 - \h\delta_\mu r^2 + {\delta_\lambda\over24}r^4 \right)\right) \cdot \nonumber\\
& & \int \mathcal{D}\psi_1 \mathcal{D}\psi_2 \; \exp\Bigg(-{1\over\hbar}\Bigg[ \h\left(\nabla\psi_1\right)^2 + \h\left(\nabla\psi_2\right)^2 + \nonumber\\
& & \phantom{\int \mathcal{D}\psi_1 \mathcal{D}\psi_2 \; \exp\Bigg(-{1\over\hbar}\Bigg(} \left( -\h\mu+{3\over2}{\mu\over v^2}r^2 \right)\psi_1^2 + \left( -\h\mu+\h{\mu\over v^2}r^2 \right)\psi_2^2 \Bigg]\Bigg) \nonumber\\
\eqa
Notice that this generating functional depends on $\delta$, as well as on the sources $J_1$ and $J_2$. With the formula 
\bq
\int \mathcal{D}\eta \; \exp\left(-{1\over\hbar} \int d^dx \left( \h\left(\nabla\eta\right)^2 + \h M^2\eta^2 \right)\right) \sim \exp\left(-\h\Omega {1\over(2\pi)^d}\int d^dk \; \ln\left(k^2+M^2\right)\right)
\eq
one can compute $Z_\delta$ further. After some algebra one finds:
\bqa
Z_{\delta} &\sim& \exp\Bigg(-{1\over\hbar}\Omega\bigg[ V_0\big(r(\delta)\cos\delta, r(\delta)\sin\delta\big) + V_1\big(r(\delta)\cos\delta, r(\delta)\sin\delta\big) \nonumber\\
& & \phantom{\exp\Bigg(-{1\over\hbar}\Omega\bigg[} -J_1r(\delta)\cos\delta - J_2r(\delta)\sin\delta \bigg]\Bigg) \;,
\eqa
with $V_0$ the classical potential
\bq
V_0 = -\h\mu\left(\f_1^2+\f_2^2\right) + {\lambda\over24}\left(\f_1^2+\f_2^2\right)^2
\eq
and $V_1$ the 1-loop effective potential found in the canonical approach, given in (\ref{effpotN2}).

\subsubsection{Including One Minimum}

Now we can see what happens if, for some reason, we would only include the single minimum. For non-zero source this minimum is at $\delta=\beta$ and $r$ given by (\ref{r}). Notice that in this case it is correct to discard the term (\ref{discpsi2term}), because $\delta=\beta$. So in this case the generating functional is given by $Z_\beta$ and the $\f_1$- and $\f_2$-tadpole can be calculated as follows.
\bqa
\langle \f_1 \rangle(J_1,J_2) &=& {\hbar\over\Omega}\frac{\partial}{\partial J_1} \ln Z_\beta \nonumber\\
&=& \left(-\frac{\partial V_0}{\partial\f_1}-\frac{\partial V_1}{\partial\f_1}+J_1\right) \frac{\partial\f_1}{\partial J_1} + \left(-\frac{\partial V_0}{\partial\f_2}-\frac{\partial V_1}{\partial\f_2}+J_2\right) \frac{\partial\f_2}{\partial J_1} + \f_1 \nonumber\\
\langle \f_2 \rangle(J_1,J_2) &=& {\hbar\over\Omega}\frac{\partial}{\partial J_2} \ln Z_\beta \nonumber\\
&=& \left(-\frac{\partial V_0}{\partial\f_1}-\frac{\partial V_1}{\partial\f_1}+J_1\right) \frac{\partial\f_1}{\partial J_2} + \left(-\frac{\partial V_0}{\partial\f_2}-\frac{\partial V_1}{\partial\f_2}+J_2\right) \frac{\partial\f_2}{\partial J_2} + \f_2 \nonumber\\
\eqa
Using
\bqa
\frac{\partial V_{0/1}}{\partial\f_1} &=& \frac{\partial V_{0/1}}{\partial r}\cos\beta \nonumber\\
\frac{\partial V_{0/1}}{\partial\f_2} &=& \frac{\partial V_{0/1}}{\partial r}\sin\beta
\eqa
and
\bqa
\frac{\partial\f_1}{\partial J_1} &=& \frac{\partial r}{\partial J_1}\cos\beta + {r\over J}\sin^2\beta \nonumber\\
\frac{\partial\f_2}{\partial J_1} &=& \frac{\partial r}{\partial J_1}\sin\beta - {r\over J}\cos\beta\sin\beta \nonumber\\
\frac{\partial\f_1}{\partial J_2} &=& \frac{\partial r}{\partial J_2}\cos\beta - {r\over J}\cos\beta\sin\beta \nonumber\\
\frac{\partial\f_2}{\partial J_2} &=& \frac{\partial r}{\partial J_2}\sin\beta + {r\over J}\cos^2\beta
\eqa
one finds
\bqa
\langle \f_1 \rangle(J_1,J_2) &=& \f_1 - \frac{\partial V_1}{\partial J_1} \nonumber\\
\langle \f_2 \rangle(J_1,J_2) &=& \f_2 - \frac{\partial V_1}{\partial J_2} \;.
\eqa

These equations can easily be inverted, up to order $\hbar$, to obtain $J_1$ and $J_2$ as a function of $\langle \f_1 \rangle$ and $\langle \f_2 \rangle$. One finds:
\bqa
J_1(\f_1,\f_2) &=& \frac{\partial V_0}{\partial\f_1}(\f_1,\f_2) + \frac{\partial V_1}{\partial\f_1}(\f_1,\f_2) \nonumber\\
J_2(\f_1,\f_2) &=& \frac{\partial V_0}{\partial\f_2}(\f_1,\f_2) + \frac{\partial V_1}{\partial\f_2}(\f_1,\f_2)
\eqa
This can be integrated to give for the effective potential, up to order $\hbar$:
\bq
V(\f_1,\f_2) = V_0(\f_1,\f_2) + V_1(\f_1,\f_2) \;.
\eq

Indeed we see that including \emph{one} minimum in the path integral gives the canonical effective potential.

\subsubsection{Including All Minima}

Including \emph{all} minima, i.e.\ all points on the ring, means:
\bq
Z = {1\over2\pi} \int_{-\pi}^\pi d\delta \; Z_\delta \;.
\eq
This generating functional can be calculated further. If we define the function $\mathcal{R}$ as
\bq
\mathcal{R}\left(J\cos(\delta-\beta)\right) = r(\delta) \;,
\eq
with $r$ defined in (\ref{r2}), the generating functional $Z$ can be written as
\bqa
Z &\sim& \int_{-\pi}^\pi d\delta \; \exp\bigg( -{1\over\hbar}\Omega\big( V_0\left(\mathcal{R}(J\cos(\delta-\beta))\right) + V_1\left(\mathcal{R}(J\cos(\delta-\beta))\right) + \nonumber\\
& & \phantom{\int_{-\pi}^\pi d\delta \; \exp\bigg( -{1\over\hbar}\Omega\big(} -\mathcal{R}(J\cos(\delta-\beta))J\cos(\delta-\beta) \big) \bigg) \nonumber\\
&\sim& \int_0^\pi d\delta \; \exp\bigg( -{1\over\hbar}\Omega\big( V_0\left(\mathcal{R}(J\cos\delta)\right) + V_1\left(\mathcal{R}(J\cos\delta)\right) + \nonumber\\
& & \phantom{\int_{-\pi}^\pi d\delta \; \exp\bigg( -{1\over\hbar}\Omega\big(} -\mathcal{R}(J\cos\delta)J\cos\delta \big) \bigg) \label{ZCartallmin}
\eqa

For the tadpoles we find
\bqa
\langle \f_1 \rangle(J_1,J_2) &=& {\hbar\over\Omega}\frac{\partial}{\partial J_1} \ln Z = {\hbar\over\Omega}\left(\frac{\partial J}{\partial J_1}\frac{\partial}{\partial J}+\frac{\partial\beta}{\partial J_1}\frac{\partial}{\partial\beta}\right) \ln Z = {\hbar\over\Omega} \cos\beta \frac{\partial}{\partial J} \ln Z \nonumber\\
\langle \f_2 \rangle(J_1,J_2) &=& {\hbar\over\Omega}\frac{\partial}{\partial J_2} \ln Z = {\hbar\over\Omega}\left(\frac{\partial J}{\partial J_2}\frac{\partial}{\partial J}+\frac{\partial\beta}{\partial J_2}\frac{\partial}{\partial\beta}\right) \ln Z = {\hbar\over\Omega} \sin\beta \frac{\partial}{\partial J} \ln Z \nonumber\\
\eqa
\bqa
& & \sqrt{\langle \f_1 \rangle^2(J_1,J_2)+\langle \f_2 \rangle^2(J_1,J_2)} = {\hbar\over\Omega}\left|\frac{\partial}{\partial J} \ln Z\right| = \nonumber\\
& & \hspace{70pt} \frac{\int_0^\pi d\delta \left[ \mathcal{R}(J\cos\delta) - \frac{\partial V_1}{\partial r}\left(\mathcal{R}(J\cos\delta)\right) \mathcal{R}'(J\cos\delta) \right]\cos\delta \;  \exp\left( \ldots \right)}{\int_0^\pi d\delta \; \exp\left( \ldots \right)} \nonumber\\ \label{phiCartallmin}
\eqa
In this last line the argument of the exponent is the same as in (\ref{ZCartallmin}).

We see that the magnitude of the $\f$-field only depends on the magnitude of the sources $J$, as expected because of the $O(2)$-symmetry.

Now this last expression is only valid for small $J$, because we discarded the term (\ref{discpsi2term}). So we will expand our result (\ref{phiCartallmin}) also in $J$ and keep all terms up to order $\hbar$. (Remember that $J$ is also of order $\hbar$.) We find:
\bq
\sqrt{\langle \f_1 \rangle^2+\langle \f_2 \rangle^2} = \frac{\int_0^\pi d\delta \left[ v\cos\delta + {J\over m^2}\cos^2\delta + \h{\Omega\over\hbar}{vJ^2\over m^2}\cos^3\delta \right] \exp\left( {\Omega vJ\over\hbar}\cos\delta \right)}{\int_0^\pi d\delta \left[ 1 + \h{\Omega\over\hbar}{J^2\over m^2}\cos^2\delta \right] \exp\left( {\Omega vJ\over\hbar}\cos\delta \right)}
\eq
This can be calculated analytically:
\bq
\sqrt{\langle \f_1 \rangle^2+\langle \f_2 \rangle^2} = \frac{vI_1\left({\Omega vJ\over\hbar}\right) + \h{J\over m^2}\left(I_0\left({\Omega vJ\over\hbar}\right)+I_2\left({\Omega vJ\over\hbar}\right)\right) + {1\over8}{\Omega\over\hbar}{vJ^2\over m^2}\left(3I_1\left({\Omega vJ\over\hbar}\right)+I_3\left({\Omega vJ\over\hbar}\right)\right)}{I_0\left({\Omega vJ\over\hbar}\right) + {1\over4}{\Omega\over\hbar}{J^2\over m^2}\left(I_0\left({\Omega vJ\over\hbar}\right)+I_2\left({\Omega vJ\over\hbar}\right)\right)}
\eq

\begin{figure}[h]
\begin{center}
\epsfig{file=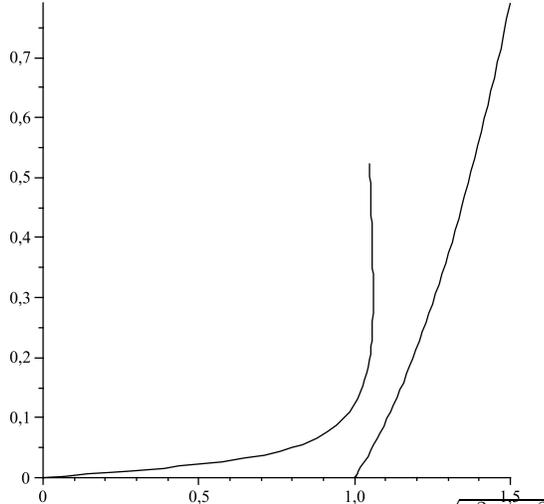,width=7cm}
\end{center}
\vspace{-1cm}
\caption{The derivative of the effective potential as a function of $\sqrt{\f_1^2+\f_2^2}/v$ for $\hbar=2$, $m=1$, $v=1$, $\Omega=100$. The left curve is only valid for small $J$, i.e. small $\sqrt{\f_1^2+\f_2^2}/v$, the right curve is only valid for large $J$.}
\label{effpotN2allmind4}
\end{figure}

This result is plotted in figure \ref{effpotN2allmind4}. The left curve is $J$, so the derivative of the effective potential, as a function of $\sqrt{\langle \f_1 \rangle^2+\langle \f_2 \rangle^2}$. The right curve is the derivative of the canonical effective potential as a function of $\sqrt{\f_1^2+\f_2^2}$. Both curves do not join at some point, the left curve is only valid for very small $J$, whereas the right curve is only valid for large $J$.

Apparently the way we calculate here, simply integrating over the ring of minima (in $r$), is \emph{not} a good way to cover the whole range of $J$, from small $J$ of order $\hbar$, to $J$ of order 1. However we \emph{do} find that the effective potential has a flat bottom in the limit $\Omega\rightarrow\infty$.

\section{The Path-Integral Approach II}\label{pathII}

In the previous section we calculated the Green's functions of the $N=2$ LSM by naively calculating Green's functions around each of the minima and then integrating over all minima. It was not at all clear that this was the correct thing to do, especially because in this approach one has to do perturbation theory around each of the minima. Each time we expand around one of these minima we pretend the ring of minima is actually an infinite line. So in this way we ignore the damping in the $\eta_2$-direction (i.e.\ tangential direction), which is there because of the $\eta_2^4$-term. This damping effect is lost in perturbation theory because the exponential of $\eta_2^4$ is expanded and not all terms are kept.

Also, by integrating over all minima we implicitly assume that the minima do not communicate, which is not true at all.

In this section we will calculate the same Green's functions via the path integral in polar field variables. These polar variables are the natural variables for a model with $O(2)$-symmetry. How one can formulate a path integral in terms of polar fields can be found in \cite{vanKessel2} and in the PhD thesis of one of the authors \cite{vanKessel}.

The action in terms of polar field variables will \emph{not} depend on the angular field $w$, but only on $\nabla w$. Therefore we have that:
\bqa
w &=& \mathcal{O}(1) \nonumber\\
\nabla w &=& \mathcal{O}(\sqrt{\hbar})
\eqa
The first relation merely states that all points on the ring of minima have an equal weight in the path integral. This means it is also incorrect to expand around $w=0$, for which we would have to assume that $w$ is small. This expansion is what we did in section \ref{pathI}. From the second relation we see that it \emph{is} correct to expand in $\nabla w$, because $\nabla w$ is small.

Because the action in terms of polar fields does not depend on $w$ there is also \emph{no} need to expand around $w=0$ in the formalism in terms of polar fields. In this way we avoid doing perturbation theory in $w$, which was the big problem of section \ref{pathI}.

In this section also the effective potential of the $N=2$ LSM will be calculated via the path integral in terms of polar fields.

\subsection{Green's Functions}

According to the conjecture from the paper \cite{vanKessel2} the path integral in terms of polar field variables for this model is given by
\bqa
& & \langle\f_1(x_1)\cdots\f_1(x_m)\f_2(y_1)\cdots\f_2(y_n)\rangle = \nonumber\\[5pt]
& & \hspace{50pt} {1\over Z(0)} \int_{-\infty}^{\infty} \mathcal{D}r \int_{-\infty}^{\infty} \mathcal{D}\theta \; \exp\left(-{1\over\hbar} I \int d^dx \left(-\hbar\ln r(x)\right)\right) \cdot \nonumber\\
& & \hspace{50pt} \phantom{{1\over Z(0)} \int_{-\infty}^{\infty} \mathcal{D}r \int_{-\infty}^{\infty} \mathcal{D}w \;} r(x_1)\cos(\theta(x_1)) \cdots r(x_m)\cos(\theta(x_m)) \cdot \nonumber\\
& & \hspace{50pt} \phantom{{1\over Z(0)} \int_{-\infty}^{\infty} \mathcal{D}r \int_{-\infty}^{\infty} \mathcal{D}w \;} r(y_1)\sin(\theta(y_1)) \cdots r(y_n)\sin(\theta(y_n)) \cdot \nonumber\\
& & \hspace{50pt} \phantom{{1\over Z(0)} \int_{-\infty}^{\infty} \mathcal{D}r \int_{-\infty}^{\infty} \mathcal{D}w \;} \exp\left(-{1\over\hbar}S(r,\theta)\right) \;,
\eqa
provided we perform the calculation in a $d$-dimensional way. Here $Z(0)$ given by
\bq
Z(0) = \int_{-\infty}^{\infty} \mathcal{D}r \int_{-\infty}^{\infty} \mathcal{D}\theta \; \exp\left(-{1\over\hbar} I \int d^dx \left(-\hbar\ln r(x)\right)\right) \exp\left(-{1\over\hbar}S(r,\theta)\right) \;,
\eq
and $S(r,\theta)$ given by
\bq
S(r,\theta) = \int d^dx \left( {1\over2}\left(\nabla r(x)\right)^2 + {1\over2}r^2(x)\left(\nabla\theta(x)\right)^2 + {\mu\over4v^2}\left( r^2(x)-v^2 \right)^2 \right) \;.
\eq

Because we are dealing with a $d$-dimensional model divergences will arise and we must renormalize the fields, masses and coupling constants. First we rewrite the action in the form
\bq \label{polaractionN2}
S(r,\theta) = \int d^dx \left( {1\over2}\left(\nabla r(x)\right)^2 + {1\over2}r^2(x)\left(\nabla\theta(x)\right)^2 - \h\mu r^2(x) + {\lambda\over24} r^4(x) \right) \;,
\eq
where
\bq
\lambda = {6\mu\over v^2} \;
\eq
The fields, masses and coupling constants are renormalized in the same way as in sections \ref{can} and \ref{pathI}:
\bqa
\varphi_i^R &=& {1\over\sqrt{Z}} \varphi_i \quad (i=1,2) \nonumber\\
\mu^R &=&\mu Z - \delta_\mu \nonumber\\
\lambda^R &=& \lambda Z^2 - \delta_{\lambda}
\eqa
In terms of polar variables the field renormalization means:
\bq
r^R = {1\over\sqrt{Z}} r \;,
\eq
the $\theta$-field is \emph{not} renormalized. We also define a new angular field as
\bq
w(x) \equiv v^R \theta(x) \;,
\eq
where
\bq
v^R = \sqrt{6\mu^R\over\lambda^R} \;.
\eq

Making these substitutions in the action (\ref{polaractionN2}) we get (also defining $\delta_Z \equiv Z-1$):
\bqa
S &=& \int d^dx \Bigg( {1\over2}\left(\nabla r^R\right)^2 + {1\over2}{\left(r^R\right)^2\over \left(v^R\right)^2}\left(\nabla w\right)^2 + {\left(\mu^R\right)^2\over4\left(v^R\right)^2}\left( \left(r^R\right)^2-\left(v^R\right)^2 \right)^2 + \nonumber\\
& & \phantom{\int d^dx \Bigg(} {1\over2} \delta_Z \left(\nabla r^R\right)^2 + {1\over2} \delta_Z {\left(r^R\right)^2\over \left(v^R\right)^2}\left(\nabla w\right)^2 - {1\over2} \delta_\mu \left(r^R\right)^2 + {1\over24} \delta_{\lambda} \left(r^R\right)^4 \Bigg) \;. \label{renormS}
\eqa

From here on we shall suppress the $R$-superscripts, understanding that we \emph{always} work with renormalized fields, masses and coupling constants.

Notice that the counter terms have nothing to do with the transformation to polar fields, both in a Cartesian and polar formulation we have the \emph{same} counter terms.

To do perturbation theory we expand around the minimum of the first line (i.e.\ the classical part) of the renormalized action:
\bq
r(x) = v + \eta(x) \;.
\eq

Remember that we also have to include the Feynman rules from the Jacobian. The procedure of renormalization does not change these rules.

The Feynman rules (in momentum space) up to order $\hbar^{5/2}$ are:
{\allowdisplaybreaks\bqa
\begin{picture}(100, 20)(0, 17)
\Line(20, 20)(80, 20)
\end{picture}
&\leftrightarrow& \frac{\hbar}{k^2 + m^2} \nonumber\\
\begin{picture}(100, 20)(0, 17)
\DashLine(20, 20)(80, 20){5}
\end{picture}
&\leftrightarrow& \frac{\hbar}{k^2} \nonumber\\
\begin{picture}(100, 40)(0, 17)
\Line(20, 20)(50, 20)
\DashLine(50, 20)(80, 0){5}
\DashLine(50, 20)(80, 40){5}
\Text(80, 10)[]{$q$}
\Text(80, 32)[]{$p$}
\end{picture}
&\leftrightarrow& \frac{2}{\hbar v} \; p\cdot q \nonumber\\
\begin{picture}(100, 40)(0, 17)
\Line(20, 0)(50, 20)
\Line(20, 40)(50, 20)
\DashLine(50, 20)(80, 0){5}
\DashLine(50, 20)(80, 40){5}
\Text(80, 10)[]{$q$}
\Text(80, 32)[]{$p$}
\end{picture}
&\leftrightarrow& \frac{2}{\hbar v^2} \; p\cdot q \nonumber\\
\begin{picture}(100, 40)(0, 17)
\Line(20, 20)(50, 20)
\Line(50, 20)(80, 0)
\Line(50, 20)(80, 40)
\end{picture}
&\leftrightarrow& -\frac{3m^2}{\hbar v} \nonumber\\
\begin{picture}(100, 40)(0, 17)
\Line(20, 40)(80, 0)
\Line(20, 0)(80, 40)
\end{picture}
&\leftrightarrow& -\frac{3m^2}{\hbar v^2} \nonumber\\
\begin{picture}(100, 40)(0, 17)
\Line(20, 20)(50, 20)
\Vertex(50,20){5}
\end{picture}
&\leftrightarrow& {v\over\hbar} \delta_\mu - {1\over6} {v^3\over\hbar} \delta_{\lambda} \nonumber\\[5pt]
\begin{picture}(100, 40)(0, 17)
\Line(20, 20)(80, 20)
\Vertex(50,20){5}
\end{picture}
&\leftrightarrow& -{1\over\hbar} \; p^2 \; \delta_Z + {1\over\hbar} \delta_\mu - {1\over2} {v^2\over\hbar} \delta_{\lambda} \nonumber\\[5pt]
\begin{picture}(100, 40)(0, 17)
\DashLine(20, 20)(80, 20){5}
\Vertex(50,20){5}
\end{picture}
&\leftrightarrow& -{1\over\hbar} \; p^2 \; \delta_Z \nonumber\\[5pt]
\begin{picture}(100, 40)(0, 17)
\Line(20, 20)(50, 20)
\DashLine(50, 20)(80, 0){5}
\DashLine(50, 20)(80, 40){5}
\Vertex(50,20){5}
\Text(80, 10)[]{$q$}
\Text(80, 32)[]{$p$}
\end{picture}
&\leftrightarrow& \frac{2}{\hbar v} \; p\cdot q \; \delta_Z \nonumber\\
\begin{picture}(100, 40)(0, 17)
\Line(20, 0)(50, 20)
\Line(20, 40)(50, 20)
\DashLine(50, 20)(80, 0){5}
\DashLine(50, 20)(80, 40){5}
\Vertex(50,20){5}
\Text(80, 10)[]{$q$}
\Text(80, 32)[]{$p$}
\end{picture}
&\leftrightarrow& \frac{2}{\hbar v^2} \; p\cdot q \; \delta_Z \nonumber\\
\begin{picture}(100, 40)(0, 17)
\Line(20, 20)(50, 20)
\Line(50, 20)(80, 0)
\Line(50, 20)(80, 40)
\Vertex(50,20){5}
\end{picture}
&\leftrightarrow& -\frac{v}{\hbar} \delta_{\lambda} \nonumber\\
\begin{picture}(100, 40)(0, 17)
\Line(20, 40)(80, 0)
\Line(20, 0)(80, 40)
\Vertex(50,20){5}
\end{picture}
&\leftrightarrow& -\frac{1}{\hbar} \delta_{\lambda} \nonumber\\
\begin{picture}(100, 40)(0, 17)
\Line(20, 20)(50, 20)
\Vertex(50, 20){3}
\end{picture}
&\leftrightarrow& {1\over v} \; I \nonumber\\
\begin{picture}(100, 40)(0, 17)
\Line(20, 20)(80, 20)
\Vertex(50, 20){3}
\end{picture}
&\leftrightarrow& -{1\over v^2} \; I \nonumber\\
\begin{picture}(100, 40)(0, 17)
\Line(20, 20)(50, 20)
\Line(50, 20)(80, 40)
\Line(50, 20)(80, 0)
\Vertex(50, 20){3}
\end{picture}
&\leftrightarrow& {2\over v^3} \; I
\eqa}
\\[10pt]
Here we have defined $\mu=\h m^2$ as in sections \ref{can} and \ref{pathI}. Also all indicated momenta flow into the vertex. The counter-term vertices have been indicated by a big dot in the vertex, the vertices from the Jacobian have been indicated by a small dot.

\subsubsection{$\eta$- And $w$-Green's-Functions} \label{O2etaw}

Now we can compute all the $\eta$- and $w$-Green's-functions.
\bqa
\langle \tilde{\eta} \rangle |_{\hbar} &=& \quad
\begin{picture}(50,40)(0,18)
\Line(0,20)(20,20)
\DashCArc(35,20)(15,-180,180){3}
\end{picture} \quad + \quad
\begin{picture}(50,40)(0,18)
\Line(0,20)(20,20)
\CArc(35,20)(15,-180,180)
\end{picture} \quad + \quad
\begin{picture}(50,40)(0,18)
\Line(0,20)(35,20)
\Vertex(35,20){5}
\end{picture} \quad + \quad
\begin{picture}(50,40)(0,18)
\Line(0,20)(35,20)
\Vertex(35,20){3}
\end{picture} \nonumber\\[25pt]
&=& -{3\over2} {\hbar\over v} \; I(0,m) + {v\over m^2} \; \delta_\mu|_{\hbar} - {1\over6} {v^3\over m^2} \; \delta_{\lambda}|_{\hbar}
\eqa
{\allowdisplaybreaks\bqa
\langle \tilde{\eta}(p) \tilde{\eta}(q) \rangle_c |_{\hbar^2} &=& \quad
\begin{picture}(70,40)(0,18)
\Line(0,20)(20,20)
\DashCArc(35,20)(15,-180,180){3}
\Line(50,20)(70,20)
\end{picture} \quad + \quad
\begin{picture}(70,40)(0,18)
\Line(0,20)(20,20)
\CArc(35,20)(15,-180,180)
\Line(50,20)(70,20)
\end{picture} \quad + \quad
\begin{picture}(70,40)(0,18)
\Line(0,5)(70,5)
\Line(35,5)(35,15)
\GCirc(35,25){10}{0.7}
\end{picture} \quad + \nonumber\\
& & \quad \begin{picture}(70,40)(0,18)
\Line(0,5)(70,5)
\DashCArc(35,20)(15,-90,270){3}
\end{picture} \quad + \quad
\begin{picture}(70,40)(0,18)
\Line(0,5)(70,5)
\CArc(35,20)(15,-90,270)
\end{picture} \quad + \quad
\begin{picture}(70,40)(0,18)
\Line(0,20)(70,20)
\Vertex(35,20){5}
\end{picture} \quad + \nonumber\\
& & \quad \begin{picture}(70,40)(0,18)
\Line(0,20)(70,20)
\Vertex(35,20){3}
\end{picture} \nonumber\\[30pt]
&=& -{\hbar^2\over v^2} \frac{p^2}{(p^2+m^2)^2} \; I(0,0) + \h {\hbar^2\over v^2} \frac{(p^2)^2}{(p^2+m^2)^2} \; I(0,0,p,0) \nonumber\\
& & +3{\hbar^2m^2\over v^2} \frac{1}{(p^2+m^2)^2} \; I(0,m) + {9\over2} {\hbar^2m^4\over v^2} \frac{1}{(p^2+m^2)^2} \; I(0,m,p,m) \nonumber\\
& & -\hbar \frac{p^2}{(p^2+m^2)^2} \; \delta_Z|_{\hbar} - 2\hbar \frac{1}{(p^2+m^2)^2} \; \delta_\mu|_{\hbar}
\eqa}
\bqa
\langle \tilde{w}(p) \tilde{w}(q) \rangle_c |_{\hbar^2} &=& \quad
\begin{picture}(70,40)(0,18)
\DashLine(0,20)(20,20){3}
\DashCArc(35,20)(15,-180,0){3}
\CArc(35,20)(15,0,180)
\DashLine(50,20)(70,20){3}
\end{picture} \quad + \quad
\begin{picture}(70,40)(0,18)
\DashLine(0,5)(70,5){3}
\Line(35,5)(35,15)
\GCirc(35,25){10}{0.7}
\end{picture} \quad + \quad
\begin{picture}(70,40)(0,18)
\DashLine(0,5)(70,5){3}
\CArc(35,20)(15,-90,270)
\end{picture} \quad + \nonumber\\
& & \quad \begin{picture}(70,40)(0,18)
\DashLine(0,20)(70,20){3}
\Vertex(35,20){5}
\end{picture} \nonumber\\[30pt]
&=& -{\hbar^2\over v^2} \frac{p^2+m^2}{(p^2)^2} \; I(0,0) + {\hbar^2\over v^2} \frac{5p^2+m^2}{(p^2)^2} \; I(0,m) + \nonumber\\
& & {\hbar^2\over v^2} \frac{(p^2+m^2)^2}{(p^2)^2} \; I(0,0,p,m) \nonumber\\
& & -\hbar \frac{1}{p^2} \; \delta_Z|_{\hbar} - 2{\hbar\over m^2} \frac{1}{p^2} \; \delta_\mu|_{\hbar} + {1\over3}{\hbar v^2\over m^2} \frac{1}{p^2} \; \delta_{\lambda}|_{\hbar}
\eqa
{\allowdisplaybreaks\bqa
\langle \tilde{\eta} \rangle|_{\hbar^2} &=& \quad
\begin{picture}(70,40)(0,18)
\Line(0,20)(20,20)
\GCirc(55,20){15}{0.7}
\DashCArc(37,20)(17,50,180){3}
\DashCArc(37,20)(17,180,310){3}
\end{picture} \quad + \quad
\begin{picture}(70,40)(0,18)
\Line(0,20)(20,20)
\GCirc(55,20){15}{0.7}
\CArc(37,20)(17,50,180)
\CArc(37,20)(17,180,310)
\end{picture} \quad + \quad
\quad \begin{picture}(70,40)(0,18)
\Line(0,20)(40,20)
\GCirc(55,20){15}{0.7}
\CArc(20,30)(10,-90,270)
\end{picture} \quad + \nonumber\\[10pt]
& & \quad \begin{picture}(70,40)(0,18)
\Line(0,20)(40,20)
\GCirc(55,20){15}{0.7}
\DashCArc(20,30)(10,-90,270){3}
\end{picture} \quad + \quad
\begin{picture}(70,40)(0,18)
\Line(0,20)(25,20)
\Line(25,20)(60,33)
\Line(25,20)(60,7)
\GCirc(60,33){10}{0.7}
\GCirc(60,7){10}{0.7}
\end{picture} \quad + \quad
\begin{picture}(70,40)(0,18)
\Line(0,20)(70,20)
\DashCArc(50,20)(20,-180,180){3}
\end{picture} \quad + \nonumber\\[10pt]
& & \quad \begin{picture}(70,40)(0,18)
\Line(0,20)(70,20)
\CArc(50,20)(20,-180,180)
\end{picture} \quad + \quad
\begin{picture}(70,40)(0,18)
\Line(0,20)(50,20)
\Vertex(50,20){5}
\end{picture} \quad + \quad
\begin{picture}(70,40)(0,18)
\Line(0,20)(40,20)
\GCirc(55,20){15}{0.7}
\Vertex(20,20){5}
\end{picture} \quad + \nonumber\\[10pt]
& & \quad \begin{picture}(70,40)(0,18)
\Line(0,20)(40,20)
\CArc(55,20)(15,0,360)
\Vertex(40,20){5}
\end{picture} \quad + \quad
\begin{picture}(70,40)(0,18)
\Line(0,20)(40,20)
\DashCArc(55,20)(15,-180,180){3}
\Vertex(40,20){5}
\end{picture} \quad + \quad
\begin{picture}(70,40)(0,18)
\Line(0,20)(40,20)
\CArc(55,20)(15,-180,180)
\Vertex(40,20){3}
\end{picture} \quad \nonumber\\[30pt]
&=& -{1\over4} {\hbar^2\over v^3} \; I(0,0)^2 + \h {\hbar^2\over v^3} \; I(0,0) I(0,m) - {3\over2} {\hbar^2m^2\over v^3} \; I(0,0) I(0,m,0,m) + \nonumber\\
& & {\hbar^2m^2\over v^3} \; D_{00m} + {3\over2} {\hbar^2m^2\over v^3} \; D_{mmm} - {3\over4} {\hbar^2m^4\over v^3} \; B_{m00} - {27\over4} {\hbar^2m^4\over v^3} \; B_{mmm} + \nonumber\\
& & -{9\over2} {\hbar^2m^2\over v^3} \; I(0,m) I(0,m,0,m) - {9\over8} {\hbar^2\over v^3} \; I(0,m)^2 + \nonumber\\
& & +{3\over2} {\hbar\over v} \; I(0,m) \; \delta_Z|_{\hbar} + {3\over2} {\hbar\over m^2v} \; I(0,m) \; \delta_\mu|_{\hbar} - {1\over4} {\hbar v\over m^2} \; I(0,m) \; \delta_{\lambda}|_{\hbar} \nonumber\\
& & -{3\over2} {\hbar m^2\over v} \; I(0,m,0,m) \; \delta_Z|_{\hbar} + 3 {\hbar\over v} \; I(0,m,0,m) \; \delta_\mu|_{\hbar} + \nonumber\\
& & -\h {v\over m^4} \; \left(\delta_\mu|_{\hbar}\right)^2 + {1\over24} {v^5\over m^4} \; \left(\delta_{\lambda}|_{\hbar}\right)^2 - {1\over6} {v^3\over m^4} \; \delta_\mu|_{\hbar} \delta_{\lambda}|_{\hbar} + {v\over m^2} \; \delta_\mu|_{\hbar^2} - {1\over6} {v^3\over m^2} \; \delta_{\lambda}|_{\hbar^2} \nonumber\\
\eqa}
\bqa
\langle \tilde{\eta}(p) \tilde{w}(q_1) \tilde{w}(q_2) \rangle_{\mathrm{c}} |_{\hbar^2} &=& \quad
\begin{picture}(70,40)(0,18)
\Line(0,20)(35,20)
\DashLine(35,20)(70,35){3}
\DashLine(35,20)(70,5){3}
\end{picture} \nonumber\\[20pt]
&=& 2{\hbar^2\over v} {1\over p^2+m^2} {1\over q_1^2} {1\over q_2^2} \; q_1\cdot q_2 \; (2\pi)^d \delta^d(p+q_1+q_2)
\eqa

\subsubsection{The $\f$-Green's-Functions}

The path integral for the $\f_1$-vacuum-expectation value is given by:
\bqa
\langle\f_1(x)\rangle &=& {1\over Z(0)} \int_{-\infty}^{\infty} \mathcal{D}r \int_{-\infty}^{\infty} \mathcal{D}w \; \exp\left(-{1\over\hbar} I \int d^dx \left(-\hbar\ln r(x)\right)\right) \cdot \nonumber\\
& & \phantom{{1\over Z(0)} \int_{-\infty}^{\infty} \mathcal{D}r \int_{-\infty}^{\infty} \mathcal{D}w \;} r(x)\cos(w(x)/v) \exp\left(-{1\over\hbar}S(r,w)\right) \;,
\eqa
where $S$ is given by (\ref{renormS}). This action \emph{only} depends on $\nabla w$, which reflects the $O(2)$-invariance of the $N=2$ LSM. This means that if we shift \emph{all} $w$-fields (i.e.\ the $w$-fields at \emph{all} space-time points) by the same amount the action does \emph{not} change. Also the path-integral measure does not change. So we can show:
\bq
\langle r(x)\cos(w(x)/v)\rangle = \langle r(x)\cos(w(x)/v+\pi)\rangle = -\langle r(x)\cos(w(x)/v)\rangle \;,
\eq
such that
\bq
\langle\f_1(x)\rangle = 0 \;.
\eq
For the same reason we have that
\bq
\langle\f_2(x)\rangle = 0 \;.
\eq
Notice that we have been able to show this through a non-perturbative argument. This is the great merit of a calculation via the path integral in terms of polar fields.

The $\f_1$- and $\f_2$-propagator can be calculated in a similar way:
\bqa
\langle\f_1(0)\f_1(x)\rangle &=& {1\over Z(0)} \int_{-\infty}^{\infty} \mathcal{D}r \int_{-\infty}^{\infty} \mathcal{D}w \; \exp\left(-{1\over\hbar} I \int d^dx \left(-\hbar\ln r(x)\right)\right) \cdot \nonumber\\
& & \phantom{{1\over Z(0)} \int_{-\infty}^{\infty} \mathcal{D}r \int_{-\infty}^{\infty} \mathcal{D}w \;} r(0)r(x) \cos(w(0)/v)\cos(w(x)/v) \cdot \nonumber\\
& & \phantom{{1\over Z(0)} \int_{-\infty}^{\infty} \mathcal{D}r \int_{-\infty}^{\infty} \mathcal{D}w \;} \exp\left(-{1\over\hbar}S(r,w)\right) \nonumber\\
&=& {1\over Z(0)} \int_{-\infty}^{\infty} \mathcal{D}r \int_{-\infty}^{\infty} \mathcal{D}w \; \exp\left(-{1\over\hbar} I \int d^dx \left(-\hbar\ln r(x)\right)\right) \; r(0)r(x) \cdot \nonumber\\
& & \phantom{{1\over Z(0)} \int_{-\infty}^{\infty} \mathcal{D}r \int_{-\infty}^{\infty} \mathcal{D}w \;} \Bigg( \h\cos\left(\frac{w(0)-w(x)}{v}\right) + \nonumber\\
& & \phantom{{1\over Z(0)} \int_{-\infty}^{\infty} \mathcal{D}r \int_{-\infty}^{\infty} \mathcal{D}w \; \Bigg(} \h\cos\left(\frac{w(0)+w(x)}{v}\right) \Bigg) \cdot \nonumber\\
& & \phantom{{1\over Z(0)} \int_{-\infty}^{\infty} \mathcal{D}r \int_{-\infty}^{\infty} \mathcal{D}w \;} \exp\left(-{1\over\hbar}S(r,w)\right) \nonumber\\
&=& \h{1\over Z(0)} \int_{-\infty}^{\infty} \mathcal{D}r \int_{-\infty}^{\infty} \mathcal{D}w \; \exp\left(-{1\over\hbar} I \int d^dx \left(-\hbar\ln r(x)\right)\right) \cdot \nonumber\\
& & \phantom{\h{1\over Z(0)} \int_{-\infty}^{\infty} \mathcal{D}r \int_{-\infty}^{\infty} \mathcal{D}w \;} r(0)r(x) \; \cos\left(\frac{w(0)-w(x)}{v}\right) \cdot \nonumber\\
& & \phantom{\h{1\over Z(0)} \int_{-\infty}^{\infty} \mathcal{D}r \int_{-\infty}^{\infty} \mathcal{D}w \;} \exp\left(-{1\over\hbar}S(r,w)\right) \nonumber\\
&=& \h \left\langle r(0)r(x) \; \cos\left(\frac{w(0)-w(x)}{v}\right) \right\rangle \nonumber\\
\langle\f_2(0)\f_2(x)\rangle &=& \langle\f_1(0)\f_1(x)\rangle \nonumber\\
\langle\f_1(0)\f_2(x)\rangle &=& 0
\eqa
Here we could discard the cosine of the sum $w(0)+w(x)$ because this cosine is \emph{not} invariant under a global shift of the $w$-field, i.e.\ this cosine is not $O(2)$-invariant.

The cosine of the difference of two $w$-fields can now be expanded, because a difference of two $w$'s can always be written as an integral over $\nabla w$, which \emph{is} small. (Remember $\nabla w=\mathcal{O}(\sqrt{\hbar})$.) 

Then the $\eta$- and $w$-Green's-functions we have calculated in the previous section can be used to find:
{\allowdisplaybreaks\bqa
\langle \varphi_1(0) \varphi_1(x) \rangle &=& {1\over2} v^2 - {1\over2} \langle w^2 \rangle + {1\over2} \langle w(0) w(x) \rangle + {1\over24v^2} \langle w^4 \rangle + {3\over24v^2} \langle w^2(0) w^2(x) \rangle + \nonumber\\
& & -{1\over12v^2} \langle w^3(0) w(x) \rangle - {1\over12v^2} \langle w(0) w^3(x) \rangle + v \langle \eta \rangle - {1\over2v} \langle \eta w^2 \rangle + \nonumber\\
& & -{1\over4v} \langle \eta(0) w^2(x) \rangle - {1\over4v} \langle \eta(x) w^2(0) \rangle + {1\over2v} \langle \eta(0) w(0) w(x) \rangle + \nonumber\\
& & {1\over2v} \langle \eta(x) w(0) w(x) \rangle + {1\over2} \langle \eta(0) \eta(x) \rangle - {1\over4v^2} \langle \eta(0) \eta(x) w^2(0) \rangle + \nonumber\\
& & -{1\over4v^2} \langle \eta(0) \eta(x) w^2(x) \rangle + {1\over2v^2} \langle \eta(0) \eta(x) w(0) w(x) \rangle + \mathcal{O}\left(\hbar^3\right) \nonumber\\[5pt]
&=& +\h v^2 \nonumber\\
& & -\h \hbar \; I(0,0) + \h \hbar \; A_0(x) - {3\over2} \hbar \; I(0,m) + \h \hbar \; A_m(x) \nonumber\\
& & +{v^2\over m^2} \; \delta_\mu|_{\hbar} - {1\over6} {v^4\over m^2} \; \delta_{\lambda}|_{\hbar} \nonumber\\
& & +\h {\hbar^2m^2\over v^2} \; I(0,0) I(0,0,0,0) - \h {\hbar^2m^2\over v^2} \; I(0,m) I(0,0,0,0) \nonumber\\
& & -{3\over2} {\hbar^2m^2\over v^2} \; I(0,0) I(0,m,0,m) - {9\over2} {\hbar^2m^2\over v^2} \; I(0,m) I(0,m,0,m) \nonumber\\
& & +\h {\hbar^2m^2\over v^2} \; D_{00m} + {3\over2} {\hbar^2m^2\over v^2} \; D_{mmm} - \h {\hbar^2m^4\over v^2} \; B_{00m} - {3\over4} {\hbar^2m^4\over v^2} \; B_{m00} \nonumber\\
& & -{27\over4} {\hbar^2m^4\over v^2} \; B_{mmm} \nonumber\\
& & -\h {\hbar^2m^2\over v^2} \; I(0,0) C_{00}(x) + \h {\hbar^2m^2\over v^2} \; I(0,m) C_{00}(x) \nonumber\\
& & +\h {\hbar^2m^2\over v^2} \; I(0,0) C_{mm}(x) + {3\over2} {\hbar^2m^2\over v^2} \; I(0,m) C_{mm}(x) \nonumber\\
& & +\h {\hbar^2m^4\over v^2} \; B_{00m}(x) + {1\over4} {\hbar^2m^4\over v^2} \; B_{m00}(x) + {9\over4} {\hbar^2m^4\over v^2} \; B_{mmm}(x) \nonumber\\
& & +\h \hbar \; I(0,0) \; \delta_Z|_{\hbar} - \h \hbar \; A_0(x) \; \delta_Z|_{\hbar} + {3\over2} \hbar \; I(0,m) \; \delta_Z|_{\hbar} - \h \hbar \; A_m(x) \; \delta_Z|_{\hbar} \nonumber\\
& & -{3\over2} \hbar m^2 \; I(0,m,0,m) \; \delta_Z|_{\hbar} + 3 \hbar \; I(0,m,0,m) \; \delta_\mu|_{\hbar} \nonumber\\
& & +\h \hbar m^2 \; C_{mm}(x) \; \delta_Z|_{\hbar} - \hbar \; C_{mm}(x) \; \delta_\mu|_{\hbar} \nonumber\\
& & +{1\over18}{v^6\over m^4} \; \left(\delta_{\lambda}|_{\hbar}\right)^2 - {1\over3}{v^4\over m^4} \; \delta_\mu|_{\hbar} \; \delta_{\lambda}|_{\hbar} + {v^2\over m^2} \; \delta_\mu|_{\hbar^2} - {1\over6}{v^4\over m^2} \; \delta_{\lambda}|_{\hbar^2} \nonumber\\[5pt]
& & +\mathcal{O}\left(\hbar^3\right) \label{symm}
\eqa}

If we compare this result to the result for the $\f_1$- and $\f_2$-propagator (\ref{phi1propchap7}), obtained in section \ref{pathI}, we find that both results agree. So, although it was far from obvious that the simple calculation done in section \ref{pathI} was correct, the result agrees with the proper calculation done in this section.

\subsubsection{Schwinger-Dyson Check}

We can check the result (\ref{symm}) by substituting it in the Schwinger-Dyson equations of the $N=2$ LSM. This check is most conveniently done on the level of the unrenormalized action.

The Schwinger-Dyson equations for the propagator can be derived through the Schwinger-Symanzik equations:
\bqa
\left[ \frac{\partial S}{\partial \varphi_1(x)}|_{\varphi_i=\hbar\frac{\partial}{\partial J_i}} - J_1(x) \right] Z\left(J_1, J_2\right) &=& 0 \nonumber\\
\left[ \frac{\partial S}{\partial \varphi_2(x)}|_{\varphi_i=\hbar\frac{\partial}{\partial J_i}} - J_2(x) \right] Z\left(J_1, J_2\right) &=& 0
\eqa
Substituting the unrenormalized action of our $N=2$ LSM, operating on both sides of the first Schwinger-Symanzik equation with $\frac{\partial}{\partial J_1(0)}$ and finally putting all sources to zero we find the Schwinger-Dyson equation for the propagator:
\bq \label{SD}
\left( \nabla^2+{1\over2}m^2 \right) \langle \varphi_1(0) \varphi_1(x) \rangle - {m^2\over2v^2} \langle \varphi_1(0) \varphi_1^3(x) \rangle - {m^2\over2v^2} \langle \varphi_1(0) \varphi_1(x) \varphi_2^2(x) \rangle = -\hbar \delta^d(x)
\eq

Now we will check the result (\ref{symm}). First we have to know the 4-points Green's functions however (not including counter terms).
\bqa
& & \langle \varphi_1(0) \varphi_1^3(x) \rangle + \langle \varphi_1(0) \varphi_1(x) \varphi_2^2(x) \rangle = \nonumber\\[10pt]
& & \hspace{30pt} \langle (v+\eta(0)) (v+\eta(x))^3 \; \cos w(0)/v \; \cos w(x)/v \rangle = \nonumber\\
& & \hspace{30pt} {1\over2} \langle (v+\eta(0)) (v+\eta(x))^3 \; \cos\left((w(0)-w(x))/v\right) \rangle = \nonumber\\
& & \hspace{30pt} {1\over2}v^4 + 2v^3 \langle \eta \rangle - {1\over2}v^2 \langle w^2 \rangle + {1\over2}v^2 \langle w(0) w(x) \rangle + {3\over2}v^2 \langle \eta^2 \rangle + {3\over2}v^2 \langle \eta(0) \eta(x) \rangle + \nonumber\\
& & \hspace{30pt} {1\over2}v \langle \eta^3 \rangle + {3\over2}v \langle \eta(0) \eta^2(x) \rangle - {3\over4}v \langle \eta(x) w^2(0) \rangle + {3\over2}v \langle \eta(x) w(0) w(x) \rangle - v \langle \eta w^2 \rangle + \nonumber\\
& & \hspace{30pt} {1\over2}v \langle \eta(0) w(0) w(x) \rangle - {1\over4}v \langle \eta(0) w^2(x) \rangle + {1\over24} \langle w^4 \rangle -{1\over12} \langle w^3(0) w(x) \rangle + \nonumber\\
& & \hspace{30pt} {1\over8} \langle w^2(0) w^2(x) \rangle - {1\over12} \langle w(0) w^3(x) \rangle - {3\over4} \langle \eta^2(x) w^2(0) \rangle + {3\over2} \langle \eta^2(x) w(0) w(x) \rangle \nonumber\\
& & \hspace{30pt} -{3\over4} \langle \eta^2 w^2 \rangle - {3\over4} \langle \eta(0) \eta(x) w^2(0) \rangle + {3\over2} \langle \eta(0) \eta(x) w(0) w(x) \rangle - {3\over4} \langle \eta(0) \eta(x) w^2(x) \rangle + \nonumber\\
& & \hspace{30pt} {1\over2} \langle \eta(0) \eta^3(x) \rangle + \mathcal{O}(\hbar^3)
\eqa
Substituting the results from section \ref{O2etaw} gives:
\bqa
& & \langle \varphi_1(0) \varphi_1^3(x) \rangle + \langle \varphi_1(0) \varphi_1(x) \varphi_2^2(x) \rangle = \nonumber\\
& & \hspace{30pt} +{1\over2}v^4 \nonumber\\
& & \hspace{30pt} -{3\over2}\hbar v^2 \; I(0,m) + {3\over2}\hbar v^2 \; A_m(x) - \h\hbar v^2 \; I(0,0) + {1\over2}\hbar v^2 \; A_0(x) \nonumber\\
& & \hspace{30pt} -\hbar^2 \; I(0,m) A_0(x) - 3\hbar^2 \; I(0,m) A_m(x) + \hbar^2 \; I(0,0) A_0(x) - \hbar^2 \; I(0,0) A_m(x) \nonumber\\
& & \hspace{30pt} +{9\over2}\hbar^2m^2 \; I(0,m) C_{mm}(x) - {9\over2}\hbar^2m^2 \; I(0,m) I(0,m,0,m) + {1\over2}\hbar^2m^2 \; I(0,m) C_{00}(x) \nonumber\\
& & \hspace{30pt} -{3\over2}\hbar^2m^2 \; I(0,0) I(0,m,0,m) + \h\hbar^2m^2 \; I(0,0) I(0,0,0,0) \nonumber\\
& & \hspace{30pt} -\h\hbar^2m^2 \; I(0,m) I(0,0,0,0) - \h\hbar^2m^2 \; I(0,0) C_{00}(x) + {3\over2}\hbar^2m^2 \; I(0,0) C_{mm}(x) \nonumber\\
& & \hspace{30pt} +{1\over2}\hbar^2m^2 \; D_{00m} - {3\over4}\hbar^2m^4 \; B_{m00} - {27\over4}\hbar^2m^4 \; B_{mmm} + {3\over2}\hbar^2m^2 \; D_{mmm} \nonumber\\
& & \hspace{30pt} -{1\over2}\hbar^2m^4 \; B_{00m} - \hbar^2m^2 \; D_{00m}(x) + {1\over2}\hbar^2m^4 \; B_{00m}(x) - {1\over2}\hbar^2m^2 \; D_{m00}(x) \nonumber\\
& & \hspace{30pt} +{3\over4}\hbar^2m^4 \; B_{m00}(x) + {27\over4}\hbar^2m^4 \; B_{mmm}(x) - {9\over2}\hbar^2m^2 \; D_{mmm}(x)
\eqa

Now, substituting this and the propagator (\ref{symm}) with all the counter terms set to zero in the Schwinger-Dyson equation (\ref{SD}) we find that the equation is satisfied.

\subsubsection{The Canonical $\f_1$-Propagator}

From our path integral in terms of polar field variables we can also recover the $\f_1$-propagator one would find in the canonical approach. To this end we have to ignore the fact that
\bq
\left\langle\cos\left(\frac{w(x)+w(y)}{v}\right)\right\rangle = 0 \;.
\eq
Instead we have to expand both cosines around $w=0$, although this is actually incorrect in the path-integral approach. Expanding the cosines around $w=0$ here corresponds to doing perturbation theory around one minimum, where we also ignore the damping in the $\eta_2$-direction and replace the ring by an infinite line. In this case we obtain:
\bqa
\langle \varphi_1(0) \varphi_1(x) \rangle_\mathrm{c} &=& \langle \varphi_1(0) \varphi_1(x) \rangle - \langle \varphi_1 \rangle^2 \nonumber\\
&=& \langle \eta(0) \eta(x) \rangle_{\mathrm{c}} - {1\over v} \langle \eta(0) w^2(x) \rangle_{\mathrm{c}} + {1\over4v^2} \langle w(0) w(x) \rangle^2 \nonumber\\
& & -{1\over v^2} \langle \eta(0) \eta(x) \rangle \langle w^2 \rangle + \mathcal{O}\left(\hbar^3\right) \nonumber\\[5pt]
&=& +\hbar \; A_m(x) \nonumber\\
& & +{\hbar^2m^2\over v^2} \; I(0,0) C_{mm}(x) + 3{\hbar^2m^2\over v^2} \; I(0,m) C_{mm}(x) + \h {\hbar^2m^4\over v^2} \; B_{m00}(x) \nonumber\\
& & +{9\over2} {\hbar^2m^4\over v^2} \; B_{mmm}(x) \nonumber\\
& & -\hbar \; A_m(x) \; \delta_Z|_{\hbar} + \hbar m^2 \; C_{mm}(x) \; \delta_Z|_{\hbar} - 2\hbar \; C_{mm}(x) \; \delta_\mu|_{\hbar} \nonumber\\[5pt]
& & +\mathcal{O}\left(\hbar^3\right) \label{phi1propcanonical}
\eqa
This propagator agrees with the $\eta_1$-propagator we found in the canonical approach (\ref{eta1propcan}).

In this $\f_1$-propagator we can substitute the counter terms (\ref{counttermsN2}) that we found in chapter \ref{can}. Then this result will satisfy the renormalization conditions from section \ref{can}.

Also this propagator can be substituted in the Schwinger-Dyson equation, together with the result for $\langle \varphi_1(0) \varphi_1^3(x) \rangle + \langle \varphi_1(0) \varphi_1(x) \varphi_2^2(x) \rangle$ in this approach, where we expand around $w=0$. These results also satisfy the Schwinger-Dyson equation. This demonstrates that \emph{both} the canonical and path-integral approach give proper solutions to the Schwinger-Dyson equations of the $N=2$ LSM.

Now we can also clearly see the \emph{difference} between results from the canonical and path-integral approach.\index{difference canonical and path-integral approach} (Compare (\ref{phi1propcanonical}) to (\ref{symm}).)

\subsection{The Effective Potential}

We can also calculate the effective potential via the path integral in terms of polar fields. To this end we introduce source terms in the renormalized action:
\bqa
S &=& \int d^dx \; \bigg( \h\left(\nabla\f_1\right)^2 + \h\left(\nabla\f_2\right)^2 - \h\mu\left(\f_1^2+\f_2^2\right) + {\lambda\over24}\left(\f_1^2+\f_2^2\right)^2 - J_1\f_1 - J_2\f_2 + \nonumber\\
& & \phantom{\int d^dx \; \bigg(} \h\delta_Z\left(\nabla\f_1\right)^2 + \h\delta_Z\left(\nabla\f_2\right)^2 - \h\delta_{\mu}\left(\f_1^2+\f_2^2\right) + {\delta_{\lambda}\over24}\left(\f_1^2+\f_2^2\right)^2 \bigg) \;.
\eqa

In section \ref{pathI} we already computed the effective potential for the $N=2$ LSM. However there we had to discard the term (\ref{discpsi2term}) to avoid ending up with an expression that contained infrared divergences at order $\hbar$. Also we could not find the interpolation of the effective potential between small $J$ (order $\hbar$) and $J$ of order 1 ($\hbar^0$). In this section we shall see if we can do a better job by calculating in terms of polar fields.

According to the conjecture the action in terms of polar fields is
\bqa
S &=& \int d^dx \; \bigg( \h\left(\nabla r\right)^2 + \h{r^2\over v^2}\left(\nabla w\right)^2 - \h\mu r^2 + {\lambda\over24}r^4 - J_1r\cos{w\over v} - J_2\sin{w\over v} + \nonumber\\
& & \phantom{\int d^dx \; \bigg(} \h\delta_Z\left(\nabla\f_1\right)^2 + \h\delta_Z\left(\nabla\f_2\right)^2 - \h\delta_{\mu}r^2 + {\delta_{\lambda}\over24}r^4 \bigg) \;,
\eqa
provided we calculate in a $d$-dimensional way in the continuum.

Introducing
\bqa
J_1 &\equiv& J\cos\beta \nonumber\\
J_2 &\equiv& J\sin\beta
\eqa
the minimum of the first line of the action, i.e.\ the classical action, is given by:
\bq
r = {2v\over\sqrt{3}} \sin\left(\alpha+{\pi\over3}\right) \;, \quad w = v\beta \;,
\eq
with
\bq
{2\mu v\over3\sqrt{3}} \sin 3\alpha = J \;.
\eq

Expanding the action around the minimum,
\bqa
r(x) &=& \bar{v} + \eta(x) \;, \qquad \bar{v} \equiv {2v\over\sqrt{3}} \sin\left(\alpha+{\pi\over3}\right) \nonumber\\
w(x) &=& v\beta + \bar{w} \;,
\eqa
we find:
\bqa
S &=& \int d^dx \; \bigg( \h\left(\nabla\eta\right)^2 + \left(-\h\mu+{3\over2}{\mu\bar{v}^2\over v^2}\right)\eta^2 + \h{\bar{v}^2\over v^2}\left(\nabla\bar{w}\right)^2 + \nonumber\\
& & \phantom{\int d^dx \; \bigg(} {\bar{v}\over v^2}\eta\left(\nabla\bar{w}\right)^2 + \h{1\over v^2}\eta^2\left(\nabla\bar{w}\right)^2 + {\mu\bar{v}\over v^2}\eta^3 + {\mu\over4v^2}\eta^4 + \nonumber\\
& & \phantom{\int d^dx \; \bigg(} -J\bar{v}\cos{\bar{w}\over v} + J\eta\left(1-\cos{\bar{w}\over v}\right) - \h\mu\bar{v}^2 + {1\over4}{\mu\bar{v}^4\over v^2} + \nonumber\\
& & \phantom{\int d^dx \; \bigg(} \h\delta_Z\left(\nabla\eta\right)^2 + \h\delta_Z{\bar{v}^2\over v^2}\left(\nabla\bar{w}\right)^2 + \delta_Z{\bar{v}\over v^2}\eta\left(\nabla\bar{w}\right)^2 + \h\delta_Z{1\over v^2}\eta^2\left(\nabla\bar{w}\right)^2 + \nonumber\\
& & \phantom{\int d^dx \; \bigg(} -\h\delta_{\mu}\bar{v}^2 + {\bar{v}^4\over24}\delta_\lambda + \nonumber\\
& & \phantom{\int d^dx \; \bigg(} \left(-\delta_\mu\bar{v}+{1\over6}\bar{v}^3\delta_\lambda\right)\eta + \left(-\h\delta_\mu+{1\over4}\bar{v}^2\delta_\lambda\right)\eta^2 + {1\over6}\bar{v}\delta_\lambda\eta^3 + {1\over24}\delta_\lambda\eta^4 \bigg) \;. \nonumber\\
\eqa
According to the conjecture the generating functional is now given by:
\bq
Z(J_1,J_2) = \int \mathcal{D}\eta \int \mathcal{D}\bar{w} \; \exp\left(-{1\over\hbar} I \int d^dx \left(-\hbar\ln\left(\bar{v}+\eta\right)\right)\right) \; e^{-{1\over\hbar}S}
\eq

As can easily be seen from the action we have \emph{one} minimum for $J\neq0$, whereas we have a ring of minima for $J=0$. This means that for $J$ of order 1 it is correct to expand the cosine of $\bar{w}/v$. In that case we recover the effective potential from the canonical approach. This is expected because for large $J$ it is correct to take into account only \emph{one} minimum. For small $J$, i.e.\ $J$ of order $\hbar$ things are a bit more difficult. For such small $J$ there is strictly speaking still one minimum, but the ring is so flat that it is incorrect to ignore the other points. We know this because when $J$ becomes really zero the other points in the ring start to play an important role. What we can do is the following. We write the generating functional as:
\bqa
Z(J_1,J_2) &=& \int \mathcal{D}\eta \int \mathcal{D}\bar{w} \; \exp\left(-{1\over\hbar} I \int d^dx \left(-\hbar\ln\left(\bar{v}+\eta\right)\right)\right) \nonumber\\
& & \phantom{\int \mathcal{D}\eta \int \mathcal{D}\bar{w} \;} \exp\left({1\over\hbar}\int d^dx \; J\left(\bar{v}+\eta\right)\cos{\bar{w}\over v}\right) \; e^{-{1\over\hbar}S'}
\eqa
with $S'$ given by:
\bqa
S' &=& \int d^dx \; \bigg( \h\left(\nabla\eta\right)^2 + \left(-\h\mu+{3\over2}{\mu\bar{v}^2\over v^2}\right)\eta^2 + \h{\bar{v}^2\over v^2}\left(\nabla\bar{w}\right)^2 + \nonumber\\
& & \phantom{\int d^dx \; \bigg(} {\bar{v}\over v^2}\eta\left(\nabla\bar{w}\right)^2 + \h{1\over v^2}\eta^2\left(\nabla\bar{w}\right)^2 + {\mu\bar{v}\over v^2}\eta^3 + {\mu\over4v^2}\eta^4 + \nonumber\\
& & \phantom{\int d^dx \; \bigg(} J\eta - \h\mu\bar{v}^2 + {1\over4}{\mu\bar{v}^4\over v^2} + \nonumber\\
& & \phantom{\int d^dx \; \bigg(} \h\delta_Z\left(\nabla\eta\right)^2 + \h\delta_Z{\bar{v}^2\over v^2}\left(\nabla\bar{w}\right)^2 + \delta_Z{\bar{v}\over v^2}\eta\left(\nabla\bar{w}\right)^2 + \h\delta_Z{1\over v^2}\eta^2\left(\nabla\bar{w}\right)^2 + \nonumber\\
& & \phantom{\int d^dx \; \bigg(} -\h\delta_{\mu}\bar{v}^2 + {\bar{v}^4\over24}\delta_\lambda + \nonumber\\
& & \phantom{\int d^dx \; \bigg(} \left(-\delta_\mu\bar{v}+{1\over6}\bar{v}^3\delta_\lambda\right)\eta + \left(-\h\delta_\mu+{1\over4}\bar{v}^2\delta_\lambda\right)\eta^2 + {1\over6}\bar{v}\delta_\lambda\eta^3 + {1\over24}\delta_\lambda\eta^4 \bigg) \;. \nonumber\\
\eqa
Note that in the new action $S'$ only $\nabla\bar{w}$ occurs. Now focus on the part that we pulled out of the action:
\bqa
& & \exp\left({1\over\hbar}\int d^dx \; J\left(\bar{v}+\eta(x)\right)\cos{\bar{w}(x)\over v}\right) \nonumber\\
& & \qquad = \exp\left({1\over\hbar}\int d^dx \; Jr(x)\cos{\bar{w}(x)\over v}\right) \nonumber\\
& & \qquad = \sum_{n=0}^\infty {1\over n!} \left( {1\over\hbar}\int d^dx \; Jr(x)\cos{\bar{w}(x)\over v} \right)^n \nonumber\\
& & \qquad = \sum_{n=0}^\infty {1\over n!} {J^n\over\hbar^n} \int d^dx_1\cdots d^dx_n \; r(x_1)\cdots r(x_n) \cos{\bar{w}(x_1)\over v}\cdots\cos{\bar{w}(x_n)\over v}
\eqa
We are going to combine the cosines into a sum of single cosines. Because the action $S'$ only depends on $\nabla\bar{w}$ only the $O(2)$-invariant cosines are going to survive in the path integral. This means, when combining the cosines, all cosines with an unequal number of $+\bar{w}$'s and $-\bar{w}$'s are going to vanish under the path integral.
\bqa
& & \exp\left({1\over\hbar}\int d^dx \; J\left(\bar{v}+\eta(x)\right)\cos{\bar{w}(x)\over v}\right) \nonumber\\
& & \qquad = \sum_{n=0}^\infty {1\over(2n)!} {J^{2n}\over\hbar^{2n}} \int d^dx_1\cdots d^dx_{2n} \; r(x_1)\cdots r(x_{2n}) \cos{\bar{w}(x_1)\over v}\cdots\cos{\bar{w}(x_{2n})\over v} \nonumber\\
& & \qquad = \sum_{n=0}^\infty {1\over(2n)!} {J^{2n}\over\hbar^{2n}} \int d^dx_1\cdots d^dx_{2n} \; r(x_1)\cdots r(x_{2n}) {2n-1\choose n} {1\over2^{2n-1}} \cdot \nonumber\\
& & \hspace{50pt} \cos{1\over v}\left(\bar{w}(x_1)+\bar{w}(x_2)+\ldots+\bar{w}(x_n)-\bar{w}(x_{n+1})-\bar{w}(x_{n+2})-\ldots-\bar{w}(x_{2n})\right) \nonumber\\
& & \qquad = \sum_{n=0}^\infty {1\over(n!)^2} \left({J\over2\hbar}\right)^{2n} \int d^dx_1\cdots d^dx_{2n} \; r(x_1)\cdots r(x_{2n}) \cdot \nonumber\\
& & \hspace{50pt} \cos{1\over v}\left(\bar{w}(x_1)+\bar{w}(x_2)+\ldots+\bar{w}(x_n)-\bar{w}(x_{n+1})-\bar{w}(x_{n+2})-\ldots-\bar{w}(x_{2n})\right) \nonumber\\
\eqa

Now we can expand the cosine, because it contains only differences of two $\bar{w}$'s. Such differences can be written as an integral over $\nabla\bar{w}$. From the path integral it can be seen that $\nabla\bar{w}$ is small (of order $\sqrt{\hbar}$), such that it is indeed correct to expand the cosine. Keeping the first and second term from the expansion of the cosine we find:
\bqa
& & \exp\left({1\over\hbar}\int d^dx \; J\left(\bar{v}+\eta(x)\right)\cos{\bar{w}(x)\over v}\right) \nonumber\\
& & \qquad = \sum_{n=0}^\infty {1\over(n!)^2} \left({J\over2\hbar}\right)^{2n} \left(\int d^dx \; r(x)\right)^{2n} + \nonumber\\
& & \qquad \phantom{=} -\h\sum_{n=0}^\infty {1\over(n!)^2} \left({J\over2\hbar}\right)^{2n} \int d^dx_1\cdots d^dx_{2n} \; r(x_1)\cdots r(x_{2n}) \cdot \nonumber\\
& & \hspace{70pt} {1\over v^2}\left(\bar{w}(x_1)+\ldots+\bar{w}(x_n)-\bar{w}(x_{n+1})-\ldots-\bar{w}(x_{2n})\right)^2 \nonumber\\
& & \qquad = \sum_{n=0}^\infty {1\over(n!)^2} \left({J\over2\hbar}\right)^{2n} \left(\int d^dx \; r(x)\right)^{2n} + \nonumber\\
& & \qquad \phantom{=} -\h\sum_{n=0}^\infty {1\over(n!)^2} \left({J\over2\hbar}\right)^{2n} \int d^dx_1\cdots d^dx_{2n} \; r(x_1)\cdots r(x_{2n}) \cdot \nonumber\\
& & \hspace{70pt} {1\over v^2} n\left( \bar{w}(x_1) - \bar{w}(x_2) \right)^2 \nonumber\\
& & \qquad = I_0\left({J\over\hbar}\int d^dx \; r(x)\right) + \nonumber\\
& & \qquad \phantom{=} -{J\over4\hbar v^2} \frac{\int d^dx \int d^dy \; r(x)r(y)\left(\bar{w}(x)-\bar{w}(y)\right)^2}{\int d^dx \; r(x)} \; I_1\left({J\over\hbar}\int d^dx \; r(x)\right)
\eqa

We can see clearly here that, if $J$ is of order 1, the first and second term are of the same magnitude (both order 1). This means that when $J$ is of order 1 we need \emph{all} terms of the expansion of the cosine. This mirrors the fact that for $J$ of order 1 there is \emph{one} clear minimum and the cosine plays a crucial role in determining where this minimum is located. Keeping all the terms of the expansion of the cosine is very hard in an actual computation, so the formula above is not very convenient to find the generating functional for $J$ of order 1.

It is convenient for $J$ of order $\hbar$ however, in this case we see that the first term above is of order 1, while the second term is of order $\hbar$. This means discarding the higher order term seems to be a good approximation. Discarding these terms means the generating functional is correct up to order $\hbar$. Also discarding other terms of higher order than $\hbar$ we find for the generating functional:
\bqa
Z(J_1,J_2) &=& \int \mathcal{D}\eta \int \mathcal{D}\bar{w} \; \exp\left(-{1\over\hbar} I \int d^dx \left(-\hbar\ln\left(\bar{v}+\eta\right)\right)\right) \; e^{-{1\over\hbar}S'} \nonumber\\
& & \qquad \Bigg[ I_0\left({J\over\hbar}\int d^dx \; (\bar{v}+\eta(x))\right) + \nonumber\\
& & \phantom{\qquad \Bigg[} -{J\over4\hbar v^2} \frac{\int d^dx \int d^dy \; (\bar{v}+\eta(x))(\bar{v}+\eta(y))\left(\bar{w}(x)-\bar{w}(y)\right)^2}{\int d^dx \; (\bar{v}+\eta(x))} \cdot \nonumber\\
& & \phantom{\qquad \Bigg[} I_1\left({J\over\hbar}\int d^dx \; (\bar{v}+\eta(x))\right) \Bigg] \nonumber\\
&=& \int \mathcal{D}\eta \int \mathcal{D}\bar{w} \; \exp\left(-{1\over\hbar} I \int d^dx \left(-\hbar\ln\left(\bar{v}+\eta\right)\right)\right) \; e^{-{1\over\hbar}S'} \nonumber\\
& & \qquad \Bigg[ I_0\left({Jv\Omega\over\hbar}\right) + I_1\left({Jv\Omega\over\hbar}\right) \; {J\over\hbar}\int d^dx \left({J\over2\mu}+\eta\right) + \nonumber\\
& & \phantom{\qquad \Bigg[} \left( I_2\left({Jv\Omega\over\hbar}\right) + {\hbar\over Jv\Omega} I_1\left({Jv\Omega\over\hbar}\right) \right) \h{J^2\over\hbar^2} \int d^dx \int d^dy \; \eta(x)\eta(y) \nonumber\\
& & \phantom{\qquad \Bigg[} -I_1\left({Jv\Omega\over\hbar}\right) {J\over4\hbar v\Omega} \int d^dx \int d^dy \; \left(\bar{w}(x)-\bar{w}(y)\right)^2 \Bigg]
\eqa
Here we have also expanded $\bar{v}$,
\bq
\bar{v} = v + {J\over2\mu} + \mathcal{O}(\hbar^2) \;,
\eq
and $\Omega$ denotes the space-time volume. Also in the action $S'$ terms of higher order than $\hbar^2$ should be discarded. (There is a $1/\hbar$ in front of the action.) In the Jacobian we should discard all terms of order higher than $\hbar$.

From the formula above one could in principle calculate the generating functional, and from it the $\f_1$- and $\f_2$-expectation-value, all up to order $\hbar$. The expectation values are given by:
\bqa
\langle\f_1\rangle(J_1,J_2) &=& {\hbar\over\Omega}\frac{\partial}{\partial J_1} \ln Z(J_1,J_2) \nonumber\\
&=& {\hbar\over\Omega} \cos\beta \frac{\partial}{\partial J} \ln Z(J_1,J_2) \nonumber\\
\langle\f_2\rangle(J_1,J_2) &=& {\hbar\over\Omega}\frac{\partial}{\partial J_2} \ln Z(J_1,J_2) \nonumber\\
&=& {\hbar\over\Omega} \sin\beta \frac{\partial}{\partial J} \ln Z(J_1,J_2)
\eqa
Notice that the generating functional does not depend on the direction of the source $\beta$. From these expectation values one can then find the effective potential.

However, when calculating the generating functional one encounters infrared divergences again. The reason is the same as in section \ref{pathI}. The formula above is only valid for small $J$, whereas we saw already in the canonical approach that to avoid the infrared singularities we need \emph{all} $n$-points Green's functions. So we also need to know $Z(J_1,J_2)$ for \emph{all} $J$, which is very hard, as we saw above. In section \ref{pathI} we could find a result (up to order $\hbar$) without infrared divergences by discarding the term (\ref{discpsi2term}), which caused the infrared divergences at order $\hbar$. In the formula above it is not clear what we can do to avoid the infrared divergences.

It is however easy to find the $\f_1$- and $\f_2$-expectation-values at lowest order from the formula for $Z$ above. We find:
\bqa
\langle\f_1\rangle(J_1,J_2) &=& v\cos\beta \; \frac{I_1\left({Jv\Omega\over\hbar}\right)}{I_0\left({Jv\Omega\over\hbar}\right)} \nonumber\\
\langle\f_2\rangle(J_1,J_2) &=& v\sin\beta \; \frac{I_1\left({Jv\Omega\over\hbar}\right)}{I_0\left({Jv\Omega\over\hbar}\right)} \;,
\eqa
and
\bq
\sqrt{\langle \f_1 \rangle^2+\langle \f_2 \rangle^2} = v\frac{I_1\left({\Omega vJ\over\hbar}\right)}{I_0\left({\Omega vJ\over\hbar}\right)} \;,
\eq
which agrees with the results from section \ref{pathI}.

The important thing however, even though we have not been able to explicitly calculate the effective potential up to order $\hbar$ in this approach, or find the interpolating form of the effective potential between the cases $J=\mathcal{O}(1)$ and $J=\mathcal{O}(\hbar)$, is that the effective potential we find is flat at the origin. This means we find the Maxwell construction of the effective potential from the canonical approach. And we find a \emph{convex} effective potential, as it should be in the path integral approach.

\section{Conclusions}

The most fundamental theory of nature known at present day is the `Standard Model'. This theory agrees very well with experimental results. All particles that are predicted in the Standard Model have also been detected in experiments, except for one: the Higgs boson. The existence of this Higgs boson in the Standard Model is derived within this model via what we call `the canonical approach'.

In the canonical approach one takes a classical field theory and quantizes it by imposing certain commutation or anti-commutation relations on the fields. The particle content of the theory is found by solving the time independent Schr\"odinger equation. One can find the vacuum state, i.e.\ the lowest energy state, via this equation, and one can build a whole Fock space on this vacuum state. The time evolution of the states is governed by the time evolution operator. Via this time evolution operator one can derive the Schwinger-Dyson equations. These equations tell one about the probability amplitudes for certain physical processes.

In the Higgs sector of the Standard Model the time independent Schr\"odinger equation is too hard to actually solve. Therefore one postulates some properties of the vacuum state, inspired by the classical lowest energy state. For example, one \emph{assumes} that the vacuum expectation value of the Higgs field is non-zero, after which one can construct the Fock space. This assumption is also very important when solving the Schwinger-Dyson equations. These Schwinger-Dyson equations can be solved iteratively. In this way one obtains a perturbative series for the Green's functions of the theory. Assuming that the vacuum expectation value of the Higgs field is non-zero one finds the Green's functions of the canonical approach. This canonical approach is completely self-consistent.

Another formulation of quantum field theory is the so-called path-integral formulation. The path integral is merely a solution to the Schwinger-Dyson equations, like the perturbative series mentioned above. For ordinary theories the path-integral formulation is just another formulation of the theory, it gives the \emph{same} physical results. The Green's functions in both formulations come out to be the same.

However, in theories for which the canonical approach predicts spontaneous symmetry breaking, it appears that both formulations of the same quantum field theory do \emph{not} yield identical results. This is the central topic of this paper. We have calculated Green's functions for the Euclidean $N=2$ LSM, for which the canonical approach predicts SSB. It appeared that, indeed, the path-integral approach gives very different Green's functions than the canonical approach.

For example, the effective potential in the canonical approach is \emph{not} convex, although one can derive, via the path-integral formulation that an effective potential should \emph{always} be convex. This is known as the convexity problem. However, it is not really a problem, because the convexity is derived in the path-integral formulation of the theory. If we accept that the canonical approach and the path-integral approach are \emph{different}, then the problem is resolved.

In the case of the $N=2$ LSM we saw that the Green's functions obtained in the canonical and path-integral approach are very different. Divergences are identical in both approaches. In section \ref{pathI} we first tried a naive approach, making some questionable steps, to take into account all minima of the path integral. In section \ref{pathII} we performed a more rigorous calculation of the path integral, based on the path integral in terms of polar fields. Results appeared to be the same. Also we obtained the effective potential of the $N=2$ LSM within the path-integral approach and found it to be convex.

With all these calculations we have established that, in the case of a theory which exhibits SSB in the canonical approach, the path-integral approach gives \emph{different} Green's functions, which may indicate different physics. This brings up some interesting questions related to the Higgs sector of the Standard Model. The prediction of the Higgs particle and its interaction are all based on the \emph{canonical} approach. What if we treat the Higgs sector of the Standard Model not in the canonical way, but instead via the path integral? What would the phenomenology of such an approach be? Could we build a theory without a Higgs particle in this way, or could we explain why the Higgs particle has not been found up to now?

\appendix
\section{Standard Integrals}\label{appstandint}

Throughout this paper we have expressed all loop integrals in terms of the following standard integrals:
\bqa
& & I\left(q_1,m_1,q_2,m_2,\ldots,q_n,m_n\right) \equiv \nonumber\\
& & \qquad {1\over(2\pi)^d} \int d^dk \; \frac{1}{\left(k+q_1\right)^2+m_1^2} \frac{1}{\left(k+q_2\right)^2+m_2^2} \cdots \frac{1}{\left(k+q_n\right)^2+m_n^2}
\eqa
\bqa
D_{m_1m_2m_3} &\equiv& {1\over(2\pi)^{2d}} \int d^dk \; d^dl \; \frac{1}{k^2+m_1^2} \; \frac{1}{l^2+m_2^2} \; \frac{1}{(k-l)^2+m_3^2} \\
B_{m_1m_2m_3} &\equiv& {1\over(2\pi)^{2d}} \int d^dk \; d^dl \; \frac{1}{(k^2+m_1^2)^2} \; \frac{1}{l^2+m_2^2} \; \frac{1}{(k-l)^2+m_3^2} \\
A_m(x) &\equiv& {1\over(2\pi)^d} \int d^dk \; e^{ik\cdot x} \; \frac{1}{k^2+m^2} \\
C_{m_1m_2}(x) &\equiv& {1\over(2\pi)^d} \int d^dk \; e^{ik\cdot x} \; \frac{1}{k^2+m_1^2} \; \frac{1}{k^2+m_2^2} \\
D_{m_1m_2m_3}(x) &\equiv& {1\over(2\pi)^{2d}} \int d^dk \; d^dl \; e^{ik\cdot x} \; \frac{1}{k^2+m_1^2} \; \frac{1}{l^2+m_2^2} \; \frac{1}{(k-l)^2+m_3^2} \\
B_{m_1m_2m_3}(x) &\equiv& {1\over(2\pi)^{2d}} \int d^dk \; d^dl \; e^{ik\cdot x} \; \frac{1}{(k^2+m_1^2)^2} \; \frac{1}{l^2+m_2^2} \; \frac{1}{(k-l)^2+m_3^2} \\
I &\equiv& {1\over(2\pi)^d} \int d^dk
\eqa


\begin{thebibliography}{99}
\bibitem{Englert} F. Englert and R. Brout, Phys.Rev.Lett. 13, 321 (1964)
\bibitem{Higgs} P.W. Higgs, Phys.Lett. 12, 132 (1964)
\bibitem{Higgs2} P.W. Higgs, Phys.Rev.Lett. 13, 508 (1964)
\bibitem{Guralnik} G.S. Guralnik, C.R. Hagen and T.W.B. Kibble, Phys.Rev.Lett. 13, 585 (1964)
\bibitem{Nambu} Y. Nambu, Phys.Rev.Lett. 4, 380 (1960)
\bibitem{NambuJona-Lasinio} Y. Nambu and G. Jona-Lasinio, Phys.Rev. 122, 345 (1961); Phys.Rev. 124, 246 (1961)
\bibitem{Bernstein} J. Bernstein, Rev.Mod.Phys. 46, 7 (1974)
\bibitem{Peskin} M.E. Peskin and D.V. Schroeder, An Introduction to Quantum Field Theory, Westview Press (1995)
\bibitem{Weinberg} S. Weinberg, The Quantum Theory of Fields, Vol. II, Cambridge University Press (1996)
\bibitem{Itzykson} C. Itzykson and J. Zuber, Quantum Field Theory, McGraw-Hill (1980)
\bibitem{Symanzik} K. Symanzik, Commun.Math.Phys. 16, 48 (1970)
\bibitem{Iliopoulos} J. Iliopoulos, C. Itzykson and A. Martin, Rev.Mod.Phys. 47, 165 (1975)
\bibitem{Haymaker} R.W. Haymaker and J. Perez-Mercader, Phys.Rev.D 27, 1948 (1983)
\bibitem{vanKessel} M.T.M. van Kessel, hep-ph: 0810.1412 (PhD thesis) (2008)
\bibitem{ORaifeartaigh} L. O'Raifeartaigh and G. Parravicini, Nucl.Phys.B 111, 516 (1976)
\bibitem{Fujimoto} Y. Fujimoto, L. O'Raifeartaigh and G. Parravicini, Nucl.Phys.B 212, 268 (1983)
\bibitem{Callaway} D.J.E. Callaway and D.J. Maloof, Phys.Rev.D 27, 406 (1983)
\bibitem{Bender} C.M. Bender and F. Cooper, Nucl.Phys.B 224, 403 (1983)
\bibitem{Cooper} F. Cooper, B. Freedman, Nucl.Phys.B 239, 459 (1984)
\bibitem{Hindmarsh} M. Hindmarsh and D. Johnston, J.Phys.A 19, 141 (1986)
\bibitem{Rivers} R.J. Rivers, Z.Phys.C 22, 137 (1984)
\bibitem{Wipf} L. O'Raifeartaigh, A. Wipf and H. Yoneyama, Nucl.Phys.B 271, 653 (1986)
\bibitem{Fukuda} R. Fukuda and E. Kyriakopoulos, Nucl.Phys.B 85, 354 (1975)
\bibitem{Ringwald} A. Ringwald and C. Wetterich, Nucl.Phys.B 334, 506 (1990)
\bibitem{Branchina} V. Branchina, P. Castorina and D. Zappal\`a, Phys.Rev.D 41, 1948 (1990)
\bibitem{Wu} E.J. Weinberg and A. Wu, Phys.Rev.D 36, 2474 (1987)
\bibitem{Dannenberg} A. Dannenberg, Phys.Lett.B 202, 110 (1988)
\bibitem{Wiedemann} U.A. Wiedemann, Nucl.Phys.B 406, 808 (1993)
\bibitem{vanKessel2} E.N. Argyres, M.T.M. van Kessel, R.H.P. Kleiss and C.G. Papadopoulos, hep-th: 0901.0815 (2009)
\bibitem{Coleman} S. Coleman, Commun.Math.Phys. 31, 259 (1973)
\bibitem{Jackiw} S. Coleman, R. Jackiw and H.D. Politzer, Phys.Rev.D 10, 2491 (1974)
\end{thebibliography}
\end{document}